\documentclass[twoside,11pt, english]{article}
\usepackage[top=2cm, bottom=2cm, outer=2.5cm, inner=2.5cm]{geometry}
\usepackage[utf8]{inputenc}
\usepackage[colorlinks,citecolor=blue,linkcolor=blue,pagebackref]{hyperref}
\usepackage{amsmath}
\usepackage{physics} 
\usepackage{amsthm}
\usepackage{amsfonts}
\usepackage{amssymb}
\usepackage{graphicx}
\usepackage[dvipsnames]{xcolor}
\usepackage{mathrsfs}
\usepackage{authblk, booktabs}
\usepackage{tikz}
\usepackage[nameinlink]{cleveref}
\usetikzlibrary{calc,decorations.markings}
\usetikzlibrary{decorations.pathmorphing}

\newtheorem{remark}{Remark}[section]

\definecolor{ClaColor}{RGB}{0,0,255}

\title{Boundary Conditions and Dirac Fields on AdS$_n$}
\author{Claudio Dappiaggi$^{1,2}$\thanks{\href{mailto:claudio.dappiaggi@unipv.it}{claudio.dappiaggi@unipv.it}}, and Andrea Parpinel$^{1,2}$\thanks{\href{mailto:andrea.parpinel01@universitadipavia.it}{andrea.parpinel01@universitadipavia.it}}}

\newcommand{\beq}{\begin{equation}}
	\newcommand{\ene}{\end{equation}}

\usepackage{pifont}
\renewcommand*{\backrefalt}[4]{%
	\ifcase #1 %
	No citations%
	\or
	\ding{43}~p.~#2%
	\else
	\ding{43}~pp.~#2%
	\fi}
\usepackage{soul}

\affil{$^{1}$Dipartimento  di  Fisica,  Universit\`a degli Studi di Pavia, Via  Bassi,  6,  27100  Pavia,  Italy}  
\affil{$^{2}$Istituto Nazionale di Fisica Nucleare --  Sezione di Pavia, Via Bassi, 6, 27100 Pavia, Italy}

\begin{document}
	
	\date{}
	
	\maketitle
	
	\begin{abstract}
We study Dirac fields on AdS$_n$ in both global and Poincar\'e charts and, for each mass window, we classify the boundary conditions at conformal infinity that ensure the existence of advanced and retarded propagators. We distinguish the well-known MIT--bag class from a generalized family, thereby extending to arbitrary dimensions the procedure initiated in~\cite{blanco}. As in the scalar case, we show that suitable generalized boundary data can support bound states. In four dimensions we work out two explicit examples: (i) the MIT case, for which we construct the advanced/retarded propagators and the two-point function of the associated ground state and (ii) a representative generalized boundary condition, for which we construct the propagators and exhibit a normalizable bound state.

	\end{abstract}

\section{Introduction}\label{Sec: Intro}

Quantum field theory on curved spacetimes is by now a mature and lively area of mathematical and theoretical physics, which relies a lot on the algebraic approach, see \cite{Benini:2013fia, Fredenhagen:2014lda, advancesAQFT} for complementary reviews. AQFT represents a very versatile and successful framework which allows both to study in depth several concrete free and interacting models and to answer relevant structural questions, ranging from the formulation of a covariant and local renormalization scheme, to the implementation of local gauge invariance, and the characterization of physically admissible states, just to quote a few notable examples. 

In many analyses of the structural aspects of quantum field theory, a common hypothesis is that of global hyperbolicity of the underlying background. Without entering in the technical aspects of the definition of this class of spacetimes, see {\it e.g.} \cite[Ch. 8]{wald}, it guarantees that solutions to symmetric hyperbolic equations, such as the Klein-Gordon, the Dirac or the Proca, can be constructed from initial data assigned on a Cauchy hypersurface. As a by product one can establish the existence of a manifestly covariant quantization scheme for free field theories, see {\it e.g.} \cite{baer}. In a few words this is based on a two-step procedure. In the first, one constructs a suitable $*$-algebra of observables $\mathcal{A}$ which encodes structural properties such as dynamics and the canonical commutation or anti-commutation relations. The second input is the choice of a state $\omega$, which can be realized abstractly as a positive and normalized linear functional on $\mathcal{A}$. Subsequently, by means of the GNS theorem, one can associate to the pair $(\mathcal{A},\omega)$ a Hilbert space representation, hence recovering the standard probabilistic interpretation proper of quantum theories. In all concrete models, fixing a state $\omega$ is tantamount to assigning the $n$-point correlation functions $\omega_n$ and, under the physically motivated assumption that the state is quasi-free/Gaussian, these can all be determined starting from $\omega_2$, see {\it e.g.} \cite[Ch. 3 \& 5]{advancesAQFT}.

Yet, among the plethora of quasi-free/Gaussian states, not all are sensible. In the literature a set of physically motivated conditions have been established. Among them, the most notable being the finiteness of the quantum fluctuations of all observables, the existence of a local and covariant regularization scheme to define Wick polynomials and the guarantee that the UV behaviour of the state is the same as that of the Poincar\'e vacuum \cite{Fewster:2013lqa}. From a technical viewpoint this translates in the now widely accepted Hadamard condition, which establishes a constraint on the singular structure of the two-point correlation function $\omega_2$. On a globally hyperbolic background the existence of Hadamard states is always guaranteed thanks to a deformation argument which relies on the property that, on static spacetimes, ground states automatically abide by the Hadamard condition \cite{verch}.   

The picture changes drastically when the assumption of global hyperbolicity is dropped. Already at the classical level, an initial value formulation is not sufficient to establish the existence of a unique solution for the underlying equation of motion, even at the level of non interacting models. The temptation of dismissing such settings as physically irrelevant is unwarranted: well-known, experimentally established phenomena, such as the Casimir effect, are naturally modelled by field theories on manifolds with boundary, see, {\it e.g.}, \cite{Dappiaggi:2014gea}. In particular, among the many lessons learned by investigating these models, noteworthy is that, at a geometric level, the lack of global hyperbolicity descends from the existence in the underlying manifold $M$ of a non empty boundary $\partial M$, consisting of one or more connected components. Furthermore, the metric induced thereon is still Lorentzian, thus endowing $\partial M$ with a lower dimensional spacetime structure.  Most notably, a central example of this class of backgrounds, whose relevance in high-energy and gravitational physics cannot be underestimated, is anti-de~Sitter spacetime AdS$_n$, of dimension $n\ge2$. This is a maximally symmetric solution of the vacuum Einstein's equations with negative cosmological constant and, up to a conformal rescaling, it possesses a timelike boundary. 

On the one hand AdS$_n$ is of paramount relevance due to its key role in several models, such as the AdS/CFT correspondence and, on the other hand, being maximally symmetric, it plays the same role of Minkowski spacetime when the cosmological constant is negative. For this reason, investigating the role of quantum field theories thereon is a very important and timely question, although the first works in this direction are from the seventies \cite{Avis:1977yn}. Already at the level of scalar free field theories one can appreciate that the paradigm at the heart of AQFT does not need to be modified, although structural differences are present. First and foremost, the construction of an algebra of observables encoding the dynamics and the canonical commutation relations rests on the existence of an advanced and retarded propagator for the Klein-Gordon operator. Yet, the presence of a boundary entails that this is not unique unless suitable boundary conditions are assigned to $\partial M$. In \cite{dappiaggi1, Dappiaggi:2017wvj} using the theory of Sturm-Liouville second order differential equations, it has been shown, that working in Fourier space and within a certain range of admissible values of the mass parameter, one can individuate a one-parameter family of admissible boundary conditions, among which one can find the Dirichlet ones, which are commonly used in the literature. To each of these one can associate advanced/retarded fundamental solutions as well as the two-point correlation function of an underlying ground state, yielding non equivalent Hilbert space representations. Such procedure has been extended to cover an even larger class of boundary conditions in \cite{Dappiaggi:2021wtr}, proving ultimately the richness of free field theories on AdS spacetimes.

At first glance one might think that the procedure which works so efficiently for a scalar field could be adapted to higher spin systems. Yet, this turns out to be false if one has to deal with first order rather than second order partial differential equations, {\it e.g.}, when working with Dirac fields. In this case, simple choices such as Dirichlet boundary conditions are known not to lead to well-defined, mixed initial-boundary value problems. Hence one has to resort to different options, among which the most famous is the MIT-bag boundary condition \cite{mit-bag}. Yet a systematic classification of the admissible boundary conditions and an analysis of the associated quantum field theory is to the best of our knowledge missing.

In view of these premises, in this work we study massive Dirac fields on AdS$_n$, both in global coordinates and in the Poincar\'e patch, denoted by PAdS$_n$. In this work we address two open questions. First, for a given mass, we classify all boundary conditions at conformal infinity that make the dynamics well posed and entail the existence of advanced and retarded Green fundamental solutions, meant as distributional inverses of the Dirac operator, which are respectively past and future propagating. We organize them into two classes: the familiar MIT--bag boundary conditions, mentioned above, and a generalized family that we identify and thoroughly characterize. This classification starts from and extends to arbitrary spacetime dimensions the procedure outlined in~\cite{blanco} for $n=2$ and it is complementary to the analysis in \cite{well-posedness} on the Cauchy problem with MIT type boundary conditions. Secondly, on the quantum side, we investigate which choices lead to physically acceptable states and whether bound states may appear. In close analogy with the scalar case, we show that within the generalized family certain parameter ranges support normalizable bound modes, while the MIT case behaves regularly.

We complement our analysis discussing in detail two examples in PAdS$_4$. For the MIT--bag boundary condition we construct the advanced and retarded propagators and a two-point function for a ground state, compatible with the PAdS$_4$ isometry group and with the analysis outlined in \cite{allen}. For a representative generalized boundary condition we again construct the propagators explicitly and exhibit a normalizable bound state. This dichotomy illustrates in concrete terms that the MIT class yields a well-behaved quantum theory, whereas the generalized class can trigger qualitatively new features through boundary-supported modes.

\vskip .3cm 

The paper is organized as follows. In Section~\ref{Sec: Geometry} we review the geometry of global AdS$_n$ and of the Poincar\'e patch, setting notation and recalling the relevant spin structure and Clifford conventions. In Section~\ref{Sec: Dirac} we formulate the Dirac equation as a first-order system on a manifold with a timelike boundary, reduce the dynamics to Fourier modes, and analyze self-adjointness of the ensuing spatial operator. In Section~\ref{sec: boundary conditions for the dirac eq} we deduce our mass-dependent classification of boundary conditions, highlighting the distinction between the MIT--bag class and its generalized counterpart. In Section \ref{sec: green operator dirac equation} we establish a prescription to construct the associated advanced and retarded fundamental solutions. In Section~\ref{Sec: Ground States} we work out two explicit four-dimensional examples: MIT and a generalized boundary condition with a normalizable bound mode. We complement our analysis with a short appendix on the Dirac equation and on our choice of gamma matrices.

\section{The Geometry of an AdS spacetime}\label{Sec: Geometry}

In this section we outline the basic geometric data at the heart of our paper. In particular, we consider the $n$-dimensional anti-de Sitter AdS$_n$, $n\geq 2$, which is a maximally symmetric solution to the Einstein's equations with a cosmological constant $\Lambda<0$. If $n>2$, it can be realized as an homogeneous, embedded submanifold of $\mathbb{R}^{n+1}$ endowed with the line element
	\begin{equation}
		\label{eq: minkowski metric for ads}
		ds^2 = -(dX^0)^2-(dX^1)^2 + (dX^2)^2 +\cdots+(dX^n)^2,
	\end{equation}
where $\{X^i\}_{i=0,\dots,n}$ are standard Cartesian coordinates. Herein the \textbf{n-dimensional anti-de Sitter spacetime} AdS$_n$ is the hyperboloid 
	\begin{equation}
		\label{eq: ads hyperboloid}
		-(X^0)^2-(X^1)^2+(X^2)^2 +\cdots+(X^n)^2 = -\ell^2 \ ,
	\end{equation}
	where $\ell^2=-\frac{(n-1)(n-2)}{\Lambda}$ is the AdS$_n$ radius of curvature. Henceforth, for simplicity we shall set $\ell=1$. In the following we shall consider both the \textit{AdS global patch} and the \textit{Poincar\'e fundamental domain} which we discuss succinctly. The first one descends from the following parameterization of Equation \eqref{eq: ads hyperboloid}:
\begin{equation}
	\label{Eq: gads coordinates}
	\begin{cases}
		X^0 = \sec{\rho}\cos{\tau}, \\
		X^1 = \sec{\rho}\sin{\tau},\\
		X^i = y^i\tan{\rho},  \quad i=2,\dots,n,
	\end{cases}
\end{equation}
where $\tau\in[-\pi,\pi)$, $\rho\in(0, \pi/2)$ while $\sum_{i=1}^{n-1}(y^i)^2 = 1$. These new coordinates cover the full AdS manifold and, inserting them in Equation \eqref{eq: minkowski metric for ads}, one obtains
\begin{equation}
	\label{Eq: gads line element}
	ds^2 = \sec^2{\rho}(-d\tau^2 + d\rho^2 + \sin^2\rho \,ds^2_{\mathbb{S}^{n-2}}), \quad i=1,\dots,n-2,
\end{equation}
where $ds^2_{\mathbb{S}^{n-2}}$ is the standard line element of the unit $(n-2)$-sphere. In order to distinguish between the different coordinate charts used in this work, we shall denote by GAdS$_n$, the manifold when decorated with the global coordinate patch as per Equation \eqref{Eq: gads coordinates}. Two comments are in due course. On the one hand we observe that GAdS$_n$, endowed with the line element as per Equation \eqref{Eq: gads line element}, is conformal to $\widetilde{\mathcal M}_n\equiv\mathbb{R}\times\mathbb{S}_+^{n-1}$, an open subset of the $n$-dimensional Einstein Static Universe ESU$_n$. Here $\mathbb{S}^{n-1}_+$ denotes half of the unit $(n-1)$-sphere whose metric is parameterized by $d\rho^2 + \sin^2\rho \,ds^2_{\mathbb{S}^{n-2}}$ with $\rho\in(0, \pi/2)$. On the other hand, in Equation \eqref{Eq: gads line element} we are making close contact with the conventions used in \cite{blanco} since, in the literature, the coordinate $\rho$ is often replaced by $r\in(0, + \infty)$, implicitly defined by $\tan{\rho} = \sinh{r}$. In addition, one can notice starting from Equation \eqref{Eq: gads coordinates} that the coordinate $\tau$ is periodic and, in order to avoid any potential causal pathology, we shall always extend the domain of $\tau$ so that $\tau\in\mathbb R$. In other words, we consider the universal cover of the $n$-dimensional anti-de Sitter spacetime which will still be denoted by GAdS$_n$ with a slight abuse of notation.

As mentioned above, we will also be considering the Poincar\'e patch/domain which is defined in terms of the coordinates
\begin{equation}
	\label{Eq: pads coordinates}
\begin{cases}
X^0 = \dfrac{t}{z},\\[6pt]
X^i = \dfrac{x^i}{z},\quad i=1,\dots,n-2,\\[6pt]
X^{n-1} = \dfrac{1}{2z}\!\left(1 + z^2 + \delta_{ij}x^i x^j - t^2\right),\\[6pt]
X^{n}   = \dfrac{1}{2z}\!\left(1 - z^2 - \delta_{ij}x^i x^j + t^2\right).
\end{cases}
\end{equation}
where $\delta_{ij}$ is the Kronecker delta, $t,x^i\in\mathbb R$, for all $i=1,\dots,n-2$, while $z\in(0,+\infty)$. Since $X^{n-1}+X^n=\frac{1}{z}>0$, the chart \eqref{Eq: pads coordinates} covers only half of AdS$_n$ and, from Equation \eqref{eq: minkowski metric for ads}, it descends that the associated line element reads
\begin{equation}\label{Eq: pads line element}
	ds^2 = \frac{1}{z^2}(-dt^2 + dz^2 + \delta_{ij}dx^i dx^j), \quad i=1,\dots,n-2.
\end{equation}
We shall denote this by PAdS$_n$, the Poincar\'e patch of AdS$_n$ and, since $z>0$, it is conformal to the upper-half space $\mathring{\mathbb H}^n:=\{(t,z,x^1,\dots,x^{n-2})\;|\;z>0\}$, endowed with the standard Minkowski metric.

\subsection{The Dirac Equation on AdS$_n$ and its boundary conditions}\label{Sec: Dirac}

On top of both GAdS$_n$ and PAdS$_n$ we wish to consider a Dirac field, following the ideas outlined in Appendix \ref{Sec: Appendix A}. Hence, focusing for definiteness on the Poincar\'e patch, we consider a spinor $\psi:\textrm{PAdS}_n\to\mathbb{C}^{2^{\lfloor\frac{n}{2}\rfloor}}$ which abides by the massive Dirac equation
\begin{equation}\label{Eq: Massive Dirac Equation}
	P \psi :=(D-m)\psi=0,
\end{equation}
where $D$ is as per Equation \eqref{Eq: dirac operator} while $m\geq 0$ is a mass parameter. As mentioned in the introduction, neither PAdS$_n$ nor GAdS$_n$ is a globally hyperbolic spacetime since they possess a timelike, conformal boundary. Hence solutions of Equation \eqref{Eq: Massive Dirac Equation} cannot be uniquely obtained only by assigning initial data on a Cauchy surface, but also suitable boundary conditions must be supplemented. In the following we classify the admissible ones extending the results of \cite{blanco} to generic spacetime dimensions. To this end we must start from an analysis of the structural aspects of the solutions of the Dirac equation both on PAdS$_n$ and on GAdS$_n$, see Section \ref{sec: pads boundary conditions}. 

\subsection{On the Dirac equation on PAdS$_n$.}\label{Sec: Solutions pads dirac eq even}

In this subsection we consider the Poincar\'e patch of the $n$-dimensional anti-de Sitter spacetime and, for computational convenience, we consider $n$ even and the following distinguished representation of the $\gamma$-matrices denoted by the symbol $\Upsilon$:  
\begin{equation}
	\label{eq: upsilon matrices}
	\Upsilon^0 = i
	\begin{pmatrix}
		0 & \mathbf{1} \\
		\mathbf{1} & 0
	\end{pmatrix},
	\quad
	\Upsilon^a = 
	\begin{pmatrix}
		-\hat{\gamma}^{a-1} & 0 \\
		0 & \hat{\gamma}^{a-1}
	\end{pmatrix},
	\quad a=1, \dots, n-1,
\end{equation}
where the matrices $\hat{\gamma}^{a-1}$ are defined in Appendix \ref{Sec: Appendix A}. Accordingly, Equation \eqref{Eq: Massive Dirac Equation} reads
\begin{equation}
	\label{Eq: Massive Dirac in PAdS}
	P\psi=(\Upsilon^\mu\nabla_\mu -m) \psi = 0.
\end{equation}
In addition, given any pair of spinors on PAdS$_n$, say $\phi,\psi:\textrm{PAdS}_n\to\mathbb{C}^{2^{\lfloor\frac{n}{2}\rfloor}}$ abiding by Equation \eqref{Eq: Massive Dirac in PAdS}, we can consider a canonical inner product subordinated to the Dirac current $J^\mu[\phi,\psi]=\bar{\phi}\Upsilon^\mu\psi$:
\begin{equation}\label{Eq: PAdSn inner product}
	\langle\phi,\psi\rangle_{\textrm{PAdS}_n}\doteq\int\limits_0^\infty\frac{dz}{z^{n-1}}\int\limits_{\mathbb{R}^{n-2}}d\underline{x}\,\bar{\phi}\Upsilon^0\psi,
\end{equation}
where $d\underline{x}=\prod_{i=1}^{n-2}dx^i$, while $\bar{\phi} = \phi^\dagger\Upsilon^0$, see Equation \eqref{eq: dirac conjugation}. Both $\phi$ and $\psi$ are here evaluated at an arbitrary but fixed value of the time coordinate $t$. This is legit since the Dirac current is covariantly conserved and therefore the right hand side of Equation \eqref{Eq: PAdSn inner product} is actually time-independent. Starting from this remark and observing also that constant time hypersurfaces on PAdS$_n$ are isometric to $\mathbb{R}^+\times\mathbb{R}^{n-2}$ endowed with the Riemannian line element $ds^2=\frac{1}{z^2}\left(dz^2+\delta_{ij}dx^i dx^j\right)$, $i,j=1,\dots,n-2$, we can introduce
\begin{equation}\label{Eq: Hilbert space PAdSn}
L^2(\mathbb{R}^+\times\mathbb{R}^{n-2};\mathbb{C}^{2^{\lfloor\frac{n}{2}\rfloor}}) = \overline{(C^\infty_0(\mathbb{R}^+\times\mathbb{R}^{n-2};\mathbb{C}^{2^{\lfloor\frac{n}{2}\rfloor}}),\langle,\rangle_{\textrm{PAdS}_n})}.
\end{equation}
This is the Hilbert space of square integrable spinors obtained as the completion of $C^\infty_0(\mathbb{R}^+\times\mathbb{R}^{n-2};\mathbb{C}^{2^{\lfloor\frac{n}{2}\rfloor}})$ with respect to the inner product \eqref{Eq: PAdSn inner product}. We highlight that the choice of $n$ even is only due to the fact, that, when $n$ is odd, the $\Upsilon$-matrices are the same as in the $(n-1)$-dimensional scenario, see Appendix \ref{Sec: Appendix A}. The following analysis should therefore be slightly modified to account for this fact, but the final result of this section applies slavishly also to $n$ odd and, therefore, for the sake of conciseness, we opted to focus only on one among the two possible scenarios. Since Equation \eqref{Eq: pads line element} entails that PAdS$_n$ is conformal to $(\mathring{\mathbb{H}}^n,\eta)$, it is convenient to consider a conformally rescaled spinor $\Psi = z^{\frac{1-n}{2}}\psi:\mathring{\mathbb{H}}^n\to\mathbb{C}^{2^{\lfloor\frac{n}{2}\rfloor}}$. In addition, if $\psi$ abides by Equation \eqref{Eq: Massive Dirac in PAdS}, then 
\begin{equation}
	\label{Eq: conformal + change basis pads dirac eq}
	(\Upsilon^\mu\nabla_\mu-m)\psi = 0 \iff \widetilde{P}\Psi=\Big(\Upsilon^\mu\partial_\mu-\frac{m}{z}\Big)\Psi = 0,
\end{equation}
where the right-hand side is the conformally related Dirac equation on $\mathring{\mathbb H}^n$. Furthermore the conformal rescaling entails that, given two solutions $\phi,\psi$ of the Dirac equation on PAdS$_n$, the inner product as per Equation \eqref{Eq: PAdSn inner product} becomes
\begin{equation}\label{Eq: Hn inner product}
	\langle\phi,\psi\rangle_{\textrm{PAdS}_n}=\int\limits_0^\infty\,dz\int\limits_{\mathbb{R}^{n-2}}d\underline{x}\,\bar{\Phi}\Upsilon^0\Psi=\langle\Phi,\Psi\rangle_{\mathring{\mathbb{H}}^n},
\end{equation}
where $\Phi=z^{\frac{1-n}{2}}\phi$, while the integral representation coincides with the natural inner product between spinors in $\mathring{\mathbb{H}}^n$. Notice that, on $\mathring{\mathbb{H}}^n$, the covariant $\Upsilon$-matrices coincide with the Minkowski ones. Following the same rationale used in Equation \eqref{Eq: Hilbert space PAdSn}, it is convenient to realize that the constant time hypersurfaces of $\mathbb{H}^n$ are isometric to $\mathbb{H}^{n-1}_\delta$ endowed with the standard $(n-1)$-dimensional Euclidean metric. Accordingly we denote by
\begin{equation}\label{Eq: Hilbert space Hn}
	L^2(\mathbb{H}^{n-1}_\delta;\mathbb{C}^{2^{\lfloor\frac{n}{2}\rfloor}}) = \overline{(C^\infty_0(\mathbb{H}^{n-1}_\delta;\mathbb{C}^{2^{\lfloor\frac{n}{2}\rfloor}}),\langle,\rangle_{\mathring{\mathbb{H}}^{n-1}})},
\end{equation}
the Hilbert space of square integrable spinors obtained as the completion of $C^\infty_0(\mathbb{H}^{n-1}_\delta;\mathbb{C}^{2^{\lfloor\frac{n}{2}\rfloor}})$ with respect to the inner product \eqref{Eq: Hn inner product}.

 In order to study Equation \eqref{Eq: conformal + change basis pads dirac eq} it is convenient to split $\mathbb{C}^{2^{\lfloor\frac{n}{2}\rfloor}}$ as $\mathbb{C}^{2^{\lfloor\frac{n}{2}\rfloor-1}}\oplus\mathbb{C}^{2^{\lfloor\frac{n}{2}\rfloor-1}}$ and consequently $\Psi =\Psi^{(1)}\oplus \Psi^{(2)}$, where $\Psi^{(1,2)}$ will be referred to as the upper- and lower-half components of the spinor $\Psi$. As a consequence, one can exploit Equation \eqref{eq: upsilon matrices} to rewrite the Dirac equation as the coupled system
\begin{subequations}
	\label{eq: pads system dirac eq}
	\begin{equation}\label{Eq: Dirac 1}
		\left(\hat{\gamma}^0\partial_z +\hat{\gamma}^j\partial_j - \frac{m}{z} \right)\Psi^{(2)} = -i\partial_t\Psi^{(1)},
	\end{equation}
	\begin{equation}\label{Eq: Dirac 2}
		\left(\hat{\gamma}^0\partial_z +\hat{\gamma}^j\partial_j + \frac{m}{z} \right)\Psi^{(1)} = i\partial_t\Psi^{(2)},
	\end{equation}
\end{subequations}
where $j=1,\dots,n-2$ labels the spatial directions orthogonal to $z$, see Equation \eqref{Eq: pads coordinates}. Deriving with respect to the time variable Equation \eqref{Eq: Dirac 1}, and employing Equation \eqref{Eq: Dirac 2}, yields 
\begin{equation}
	\label{eq: klein-gordon pads eq}
	\Big(\eta^{\mu\nu}\partial_\mu\partial_\nu  -  \hat{\gamma}^0 \frac{m}{z^2} - \frac{m^2}{z^2}  \Big)\Psi^{(1)} = 0 \ ,
\end{equation}
where we exploited that the $\gamma$-matrices are time-independent. Reversing the role of $\Psi^{(1)}$ and $\Psi^{(2)}$ yields
\begin{equation}
	\label{eq: klein-gordon pads 2 eq}
	\Big(\eta^{\mu\nu}\partial_\mu\partial_\nu  +  \hat{\gamma}^0 \frac{m}{z^2} - \frac{m^2}{z^2}  \Big)\Psi^{(2)} = 0 \ .
\end{equation}
Since the translations along all directions tangent to the boundary of $\mathring{\mathbb{H}}^n$ are isometries of Equation \eqref{Eq: pads line element}, it is convenient to assume that $\Psi^{(i)}\in\mathcal{S}^\prime(\mathring{\mathbb{H}}^n)$, $i=1,2$, so that, working at the level of integral kernel, we can take the Fourier transform along all coordinates barring $z$:
\begin{equation}
	\label{eq: pads fourier transform}
	\Psi^{(1)}(t,z,x^i) = \int_{\mathbb R^{n-1}} \frac{d\omega \,d\mathbf k}{(2\pi)^{\frac{n-1}{2}}} \, e^{i(\omega t- \mathbf{k}\cdot\mathbf{x})} \Psi^{(1)}_{\omega,\mathbf k}(z),
\end{equation}
where $\mathbf{x}=(x^1,\dots,x^{n-2})\in\mathbb R^{n-2}$, $\omega\in\mathbb R$, while $\mathbf{k} = (k^1,\dots k^{n-2})\in\mathbb R^{n-2}$. The expression $\mathbf k \cdot \mathbf x$ denotes the standard Euclidean inner product. Plugging Equation \eqref{eq: pads fourier transform} inside Equation \eqref{eq: klein-gordon pads eq} yields the following second-order, vector valued, ordinary differential equation:
\begin{equation}
	\label{eq: pads SL problem}
	\Big(\frac{d^2}{dz^2}  -  \hat{\gamma}^0 \frac{m}{z^2} - \frac{m^2}{z^2} +  \omega^2 - |\mathbf{k}|^2 \Big)\Psi^{(1)}_{\omega, \mathbf k}(z) = \Big(\frac{d^2}{dz^2}  -  \hat{\gamma}^0 \frac{\nu-\frac{1}{2}}{z^2} - \frac{(\nu-\frac{1}{2})^2}{z^2} +  \omega^2 - |\mathbf{k}|^2 \Big)\Psi^{(1)}_{\omega, \mathbf k}(z) = 0 \ ,
\end{equation}
where $z\in(0,+\infty)$, $\nu\doteq m+\frac12$, while $|\mathbf k|^2$ denotes the standard Euclidean norm of $\mathbf{k}\in\mathbb R^{n-2}$. At last, given $\hat\gamma^0$ as per Appendix \ref{Sec: Appendix A} , Equation \eqref{eq: pads SL problem} is a system of $\frac{N}{2} = 2^{\lfloor\frac{n}{2}\rfloor -1}$ decoupled, second-order ordinary differential equations of Sturm-Liouville type, see \cite{zettl}. The collection of linearly independent solutions of Equation \eqref{eq: pads SL problem} reads
\begin{align}
	\label{Eq: pads solution dirac eq}
	\begin{split}
		\mathbf{J}_i^{(\nu)}(z) &= [\sqrt{z}J_{\nu}(qz)]e_i, \quad  i=1,\dots,\frac{N}{4}, \\
		\mathbf{Y}^{(\nu)}_i(z) &= [\sqrt{z}Y_{\nu}(qz)]e_i, \quad i=1,\dots,\frac{N}{4}, \\
		\mathbf{J}_i^{(\nu-1)}(z) &= [\sqrt{z}J_{\nu-1}(qz)]e_i, \quad i=\frac{N}{4}+1,\dots,\frac{N}{2}, \\
		\mathbf{Y}^{(\nu-1)}_i(z) &= [\sqrt{z}Y_{\nu-1}(qz)]e_i, \quad i=\frac{N}{4}+1,\dots,\frac{N}{2},
	\end{split}
\end{align}
where $q^2\doteq\omega^2-|\mathbf k|^2$, $\{e_i\}_{i=1,\dots,\frac{N}{2}}$ is the canonical basis of $\mathbb R^{\frac{N}{2}}$, $J_\nu$ denotes the Bessel function of first kind of order $\nu$, while $Y_\nu$ that of second kind of the same order, see \cite{NIST:DLMF}. It is worth observing that, from a mere mathematical viewpoint, contrary to what occurs when working with a real scalar field on PAdS$_n$, there is no reason to restrict the values of $\nu$ and therefore of $m$ on account of Equation \eqref{Eq: pads solution dirac eq}. In other words, a priori, at the level of ODEs, there exists no counterpart of the Breitenlohner-Freedman bound \cite{BF, BF2}. Yet, as we can infer from direct inspection of Equation \eqref{Eq: pads solution dirac eq} the set of the first $\frac{N}{2}$ ODEs is invariant under the exchange $\nu\to-\nu$ while the second half under $\nu\to 2-\nu$. Taking into account both symmetries, we can limit ourselves to considering the case $\nu\geq 0$ as the remaining values can be studied exploiting the mentioned symmetry. Henceforth we shall always stick to this choice and, furthermore, we shall discard the extremal case $\nu=0$ since it would require a separate analysis. Focusing instead on $\Psi^{(2)}$ and working once more at the level of Fourier transform, Equation \eqref{eq: pads system dirac eq} yields
\begin{equation}
	\label{Eq: lower half spinor pads}
	\Psi^{(2)}_{\omega, \mathbf k}(z) = -\frac{1}{\omega}\Big(\hat\gamma^0\frac{d}{dz} -ik_j\hat{\gamma}^j +\frac{m}{z}\Big)\Psi^{(1)}_{\omega, \mathbf k}(z),
\end{equation}
which entails that Equation \eqref{Eq: pads solution dirac eq} gets transformed into
\begin{align}
	\label{Eq: lower half solutions dirac eq pads}
	\begin{split}
		\widetilde{\mathbf{J}}_i^{(\nu)}(z) &= -\frac{\sqrt{z}}{\omega}[qJ_{\nu-1}(qz)\mathbf{1} -iJ_{\nu}(qz)k_j\hat{\gamma}^j]e_i, \quad i=1,\dots,\frac{N}{4}, \\
		\widetilde{\mathbf{Y}}^{(\nu)}_i(z) &= -\frac{\sqrt{z}}{\omega}[qY_{\nu-1}(qz)\mathbf{1} -iY_{\nu}(qz)k_j\hat{\gamma}^j]e_i, \quad i=1,\dots,\frac{N}{4}, \\
		\widetilde{\mathbf{J}}_i^{(\nu-1)}(z) &= \frac{\sqrt{z}}{\omega}[qJ_{\nu}(qz)\mathbf{1} +iJ_{\nu-1}(qz)k_j\hat{\gamma}^j]e_i, \quad i=\frac{N}{4}+1,\dots,\frac{N}{2}, \\
		\widetilde{\mathbf{Y}}^{(\nu-1)}_i(z) &= \frac{\sqrt{z}}{\omega}[qY_{\nu}(qz)\mathbf{1} +iY_{\nu-1}(qz)k_j\hat{\gamma}^j]e_i, \quad i=\frac{N}{4}+1,\dots,\frac{N}{2}.
	\end{split}
\end{align}

\begin{remark}\label{remark: zero mode solution}
    From Equation \eqref{Eq: lower half spinor pads} it is clear that the above procedure is not valid for $\omega=0$. When considering this particular value, one has to take into account that Equation \eqref{eq: pads system dirac eq} reduces to a decoupled system of differential equations whose solutions can be expressed in terms of modified Bessel functions $K_\alpha(|\mathbf k|z), \, I_\alpha(|\mathbf k|z)$, of order $\nu$ and $\nu-1$. In Section \ref{Sec: Ground States} we show how these solutions might contribute in the construction of the propagators of the Dirac equation.
\end{remark}

\subsection{Solutions of the Dirac equation on GAdS$_n$}\label{Sec: gads solution even case}

We focus our attention on GAdS$_n$ whose associated line element is codified in Equation \eqref{Eq: gads coordinates} and, exactly as in the previous section, we set $n$ even. Since Equation \eqref{Eq: gads line element} entails that GAdS$_n$ is conformal to $\widetilde{\mathcal M}_n\subset\textrm{ESU}_n$, we consider a conformally rescaled spinor $\Psi = (\sec\rho)^{\frac{1-n}{2}}\psi:\widetilde{\mathcal M}_n\to\mathbb{C}^{2^{\lfloor\frac{n}{2}\rfloor}}$so that, if $\psi$ abides by Equation \eqref{Eq: Massive Dirac in PAdS} and if we work with the $\Upsilon$-matrices as per Equation \eqref{eq: upsilon matrices}, then 
\begin{equation}
	\label{eq: conformal dirac eq gads}
	(\Upsilon^\mu\nabla_\mu - m\sec{\rho})\Psi = 0,
\end{equation}
where $\nabla_\mu$ is the spin connection on $\widetilde{\mathcal M}$ as per Equation \eqref{eq: spinor covariant derivative} built out of the $\Upsilon$-matrices. In addition, given any pair of spinors on GAdS$_n$, say $\phi,\psi:\textrm{GAdS}_n\to\mathbb{C}^{2^{\lfloor\frac{n}{2}\rfloor}}$, both solutions of the Dirac equation, we can consider a canonical inner product subordinated to the Dirac current $J^\mu[\phi,\psi]=\bar{\phi}\Upsilon^\mu\psi$, from which we can construct the following inner product:
\begin{equation}\label{Eq: GAdSn inner product}
	\langle\phi,\psi\rangle_{\textrm{GAdS}_n}\doteq\int\limits_{0}^{\frac{\pi}{2}}d\rho\,\sec^{n-1}\rho\int\limits_{\mathbb{S}^{n-2}}d\mathbb{S}^{n-2}\,\bar{\phi}\Upsilon^0\psi,
\end{equation}
where $d\mathbb{S}^{n-2}$ is the standard volume form on the unit $(n-2)$-sphere, while $\bar{\phi}$ is the Dirac conjugate as per Equation \eqref{eq: dirac conjugation}. As in the previous section, both $\phi$ and $\psi$ are here implicitly evaluated at an arbitrary but fixed value of the time coordinate $t$. Yet, since the Dirac current is covariantly conserved this choice is irrelevant since the right hand side of Equation \eqref{Eq: GAdSn inner product} does not depend on time. Furthermore the conformal rescaling entails that the inner product as per Equation \eqref{Eq: GAdSn inner product} becomes
\begin{equation}\label{Eq: ESU inner product}
	\langle\phi,\psi\rangle_{\textrm{GAdS}_n}\doteq\int\limits_{0}^{\frac{\pi}{2}}d\rho\,\int\limits_{\mathbb{S}^{n-2}}d\mathbb{S}^{n-2}\,\bar{\Phi}\Upsilon^0\Psi=\langle\Phi,\Psi\rangle_{\widetilde{\mathcal M}_n},
\end{equation}
where $\Phi=\sec^{\frac{1-n}{2}}\rho \, \phi$, while the integral representation coincides with the natural inner product between spinors in $\widetilde{\mathcal M}_n$. As in the previous section, it is convenient to split $\Psi =\Psi^{(1)}\oplus \Psi^{(2)}$ and to write
\begin{equation}
	\label{eq: dirac operator gads even case}
	\Upsilon^\mu\nabla_\mu = \begin{pmatrix}
		-D_{S^{n-1}} & i\partial_t \\
		i\partial_t & D_{S^{n-1}},
	\end{pmatrix},
\end{equation}
where $D_{S^{n-1}}$ is the Dirac operator on $S^{n-1}$, the unit $(n-1)$-sphere. This yields the system
\begin{equation}
	\begin{cases}
		(D_{S^{n-1}} + m\sec{\rho} ) \Psi^{(1)} = i\partial_t \Psi^{(2)} \\
		(D_{S^{n-1}} - m\sec{\rho} ) \Psi^{(2)} = -i\partial_t \Psi^{(1)}
	\end{cases}.
\end{equation}
Since the Dirac operator on $\mathbb{S}^{n-1}$ is time independent, by applying to the second equation $\partial_t$ and inserting the first one, it descends that
\begin{equation}
	\label{eq: gads SL problem}
	\Big(-\frac{\partial^2}{\partial t^2} + D_{S^{n-1}}^2 + \hat\gamma^0 \, m \sec{\rho}\tan{\rho} - m^2\sec^2{\rho} \Big)\Psi^{(1)} = 0\,.
\end{equation}
As discussed thoroughly in \cite{gamma-matrices}, $D_{S^{n-1}}$ has a discrete, completely imaginary spectrum, $\lambda_k=i(k+\frac{n-1}{2})$, $k\in\mathbb{Z}$. Its eigenfunctions are the \textit{spinor spherical harmonics} $\zeta_{k,m}(\rho,\Omega^i)$, where $m=-|k|,\dots, |k|$, while $\{\Omega^i\}_{i=1,\dots,n-2}$ denote the standard angular coordinates of $S^{n-2}$. These harmonics are orthogonal with respect to the $L^2(S^{n-1})$ inner product and they form a complete basis for the space of square-integrable spinors on $S^{n-1}$. Assuming that $\Psi^{(1)}$ is a tempered distribution along the time direction we can take the Fourier transform and write 
\begin{equation}
	\label{eq: gads anstatz half spinor}
	\Psi^{(1,\pm)} (t,\rho,\Omega^i) = \sum_{k\in\mathbb{Z}}\sum_{m=-k}^{k} \int_{\mathbb R} \frac{d\omega}{\sqrt{2\pi}} \,  e^{i(\omega t \mp\lambda\rho)}\, \frac{1}{2}(\mathbf 1\pm\hat\gamma^0)f^{(1,\pm)}_{\omega,\lambda}(\rho) \, \zeta_\lambda(\rho,\Omega^i).
\end{equation}
 The superscripts $\pm$ indicate that we are splitting $\Psi^{(1)}$ further into two half-spinors. In fact, one can verify that $\frac{1}{2}(1\pm\hat\gamma^0)$ is the projector on the upper-half or lower-half components for the half-spinors $\Psi^{(1,2)}$.  The reader should be aware of the fact that, in general, the sum in Equation \eqref{eq: gads anstatz half spinor} should be regarded as a multi-indexed sum over a countable set which takes into account also the multiplicities of the eigenvalues. Combining Equations \eqref{eq: gads SL problem} and \eqref{eq: gads anstatz half spinor} we end up with the following ordinary differential equation:
\begin{equation}
	\label{eq: gads ODE}
	\Big(\frac{d^2}{d\rho^2} \pm m \sec{\rho}\tan{\rho} - m^2\sec^2{\rho} + \omega^2 \Big )f^{(1,\pm)}_{\omega,\lambda}(\rho) = 0\,.
\end{equation}
Generalizing the analysis in \cite{blanco} for $n=2$, the linearly independent solutions of Equation \eqref{eq: gads ODE} read
\begin{align}
	\label{gads solutions list}
	\begin{split}
		f_{\omega, \lambda}^{(1,\pm)}(\rho) &= (2\nu_\pm+1)\,\sigma(\rho)^{\pm m} \, F\Big(\omega,-\omega;\nu_\pm;\frac{1-\sin{\rho}}{2} \Big), \\
		h_{\omega, \lambda}^{(1,\pm)}(\rho) &= \omega\,\cos{\rho}\,\sigma(\rho)^{\mp m} \, F\Big(1+\omega,1-\omega;2-\nu_\pm;\frac{1-\sin{\rho}}{2} \Big),
	\end{split}
\end{align}
where $\nu_\pm\doteq \pm m+\frac12$, $\sigma(\rho) \doteq (\frac{1-\sin\rho}{1+\sin\rho})^{\frac12}$, while $F(a,b;c;z)$ denotes the Gaussian hypergeometric function, see \cite{NIST:DLMF}. Eventually, using Equation \eqref{gads solutions list}, one determines $\Psi^{(2)}$ replacing in Equation \eqref{eq: gads anstatz half spinor} $f^{(1,\pm)}_{\omega,\lambda}(\rho)$ with
\begin{equation}
	\label{eq: gads lower spinor}
	f^{(2,\pm)}_{\omega,\lambda} (\rho)= -\frac{1}{\omega}\Big(\pm\frac{d}{d\rho} + m\sec{\rho} \Big ) f^{(1,\pm)}_{\omega,\lambda}(\rho).
\end{equation}

To conclude the section, we highlight that the preceding analysis can be repeated also for $n$ odd, though taking into account the differences in the form of the $\Upsilon$-matrices. Hence Equations \eqref{eq: gads anstatz half spinor} and \eqref{gads solutions list} still apply. Observe also that, mutatis mutandis, the content of Remark \ref{remark: zero mode solution} also applies to this scenario.

\section{Boundary conditions for the Dirac equation}
\label{sec: boundary conditions for the dirac eq}
Having constructed the solutions of the Dirac equation both on PAdS$_n$ and on GAdS$_n$, we investigate the class of admissible boundary conditions which can be assigned to the underlying dynamical problem in addition to initial data, so to single out a unique solution. In \cite{dappiaggi1}, when discussing the same issue for a real scalar field, one had to resort to using techniques proper of Sturm-Liouville problems, see \cite{zettl,green-function-bvp}. On the contrary, since the Dirac operator identifies a first order partial differential equation, we can rely on the fact, that, both in PAdS$_n$ and in GAdS$_n$, the Dirac equation can be written in a Schr\"{o}dinger form:
\begin{equation}
	\label{Eq: dirac eq in schrodinger form}
	i\partial_t\psi  = H\psi,
\end{equation}
where, having set for simplicity $\hbar=1$, $t$ is the time coordinate while $H$ is a suitable differential operator which plays the role of an \textit{Hamiltonian}. As it will emerge from the following analysis, we shall be interested in interpreting $H$ as a symmetric operator on the space of square-integrable spinors living on constant time hypersurfaces with respect to the inner product \eqref{Eq: PAdSn inner product} on PAdS$_n$ and \eqref{Eq: GAdSn inner product} on GAdS$_n$. Most notably $H$ shall admit self-adjoint extensions, each of which will be associated to a boundary condition. From the operator theoretical viewpoint this serves as a constraint on the domain of $H$ itself while, from that of the Dirac equation \eqref{Eq: Massive Dirac Equation}, it will identify an admissible boundary condition yielding in particular a unique advanced and retarded fundamental solution. In the following we will be employing the theory of deficiency indices and we refer to \cite{Moretti:2013cma} for an introduction to this topic.

\subsection{PAdS$_n$ case}\label{sec: pads boundary conditions}
In this section we consider the Dirac equation on PAdS$_n$ as per Equation \eqref{Eq: Massive Dirac in PAdS} and, exploiting the defining properties of the $\Upsilon$-matrices as per Equation \eqref{eq: covariant gamma matrices}, we recast it in a Schr\"odinger-like form:
\begin{equation}
	\label{Eq: dirac hamiltonian pads}
	i\partial_t\psi = i\Upsilon^0(\Upsilon^k\nabla_k-m)\psi \doteq H\psi, \quad k =1,\dots,n-1,
\end{equation}
where $H\doteq i\Upsilon^0(\Upsilon^k\nabla_k-m)$ is the \textit{Dirac Hamiltonian} on PAdS$_n$. As in Section \ref{Sec: Solutions pads dirac eq even}, we can exploit the conformal rescaling $\psi\to\Psi\doteq z^{\frac{1-n}{2}}\psi$ to individuate an equivalent equation on $\mathring{\mathbb{H}}^n$, namely
\begin{equation}
	\label{Eq: conformal dirac hamiltonian pads}
	i\partial_t \Psi = i\Upsilon^0 \Big( \Upsilon^k\partial_k-\frac{m}{z} \Big) \Psi \doteq \mathsf{H} \Psi, \quad k=1,\dots,n-1,
\end{equation}
where $\mathsf{H}\doteq i\Upsilon^0 \Big( \Upsilon^k\partial_k-\frac{m}{z} \Big)$ is the {\em conformal Dirac Hamiltonian}. Notice that, with a slight abuse of notation, we are denoting by the same symbols the covariant $\Upsilon$-matrices on PAdS$_n$ and on $\mathring{\mathbb{H}}^n$, though $\Upsilon^\mu_{\mathring{\mathbb{H}}^n} = \frac{1}{z}\Upsilon^\mu_{\textrm{PAdS}_n}$. Equations \eqref{Eq: dirac hamiltonian pads} and \eqref{Eq: conformal dirac hamiltonian pads} suggest to read $H$ ({\em resp.} $\mathsf{H}$) as an operator on $L^2(\mathbb{R}^+\times\mathbb{R}^{n-2};\mathbb{C}^{2^{\lfloor\frac{n}{2}\rfloor}})$, see Equation \eqref{Eq: Hilbert space PAdSn} ({\em resp.} on $L^2(\mathbb{H}^{n-1}_\delta;\mathbb{C}^{2^{\lfloor\frac{n}{2}\rfloor}})$, see Equation \eqref{Eq: Hilbert space Hn}). Observe that $H$ is manifestly symmetric, namely
$$\langle\phi,H\psi\rangle_{\textrm{PAdS}_n}-\langle H\phi,\psi\rangle_{\textrm{PAdS}_n}=0, \quad\forall\phi,\psi\in C^\infty_0(\mathbb{R}^+\times\mathbb{R}^{n-2};\mathbb{C}^{2^{\lfloor\frac{n}{2}\rfloor}})$$ 
and a similar conclusion can be drawn for $\mathsf{H}$. Therefore it is meaningful to look for their self-adjoint extensions, hence extending the procedure used in \cite{blanco} on $GAdS_2$ and, on account of Equation \eqref{Eq: Hn inner product}, it is sufficient to discuss one among the two operators. Focusing on $\mathsf{H}$, we start by using Equation \eqref{eq: upsilon matrices} to write it in the following matrix form: 
\begin{equation*}
	\mathsf{H} = i\Upsilon^0 \Big(\Upsilon^k\partial_k - \frac{m}{z}\Big) = \begin{pmatrix}
		0 & -\hat\gamma^0\partial_z- \hat\gamma^i\partial_i+\frac{m}{z} \\
		\hat\gamma^0\partial_z+\hat\gamma^i\partial_i+\frac{m}{z} & 0 
	\end{pmatrix},
\end{equation*}
where the index $i$ runs over all coordinates barring $t$ and $z$. In analogy to Equation \eqref{eq: pads fourier transform}, since $D(\mathsf{H})$, the domain of $\mathsf{H}$ or of any of its self-adjoint extensions, must be contained in $L^2(\mathbb{H}^{n-1}_\delta;\mathbb{C}^{2^{\lfloor\frac{n}{2}\rfloor}})$, we can consider the Fourier-Plancherel transform along all directions tangent to the boundary. In other words, for any $\Psi\in D(\mathsf{H})$, it holds
\begin{equation}
	\label{eq: operator D pads}
\mathsf{H}\Psi= \int_{\mathbb R^{n-2}} \frac{d\mathbf k}{(2\pi)^{\frac{n-2}{2}}} \, e^{-i\mathbf k \cdot \mathbf x} \,\mathbb D_{\mathbf k}\Psi_{\mathbf k}(z) \, ,\quad\mathbb D_{\mathbf k} \doteq 
		\begin{pmatrix}
			0 & -\hat\gamma^0\frac{d}{dz}+i \hat\gamma^ik_i+\frac{m}{z} \\
			\hat\gamma^0\frac{d}{dz}-i\hat\gamma^ik_i+\frac{m}{z} & 0 
		\end{pmatrix},
\end{equation}
where the subscript $\mathbf k$ highlights the dependence on the momenta. As a consequence we can focus our attention on $\mathbb{D}_{\mathbf k}$ seen as an operator on $L^2((0,\infty);\mathbb{C}^{2^{\lfloor\frac{n}{2}\rfloor}})$ where, here, we consider the standard Lesbegue measure. As a first step we read $\mathbb{D}_{\mathbf k}$ as an operator with domain Dom$(\mathbb{D}_{\mathbf k})\equiv C^\infty_0((0,\infty);\mathbb{C}^{2^{\lfloor\frac{n}{2}\rfloor}})$ and, for every $\Psi_{\mathbf k}, \Phi_{\mathbf k}\in\textrm{Dom}(\mathbb{D}_{\mathbf k})$, it holds integrating by parts that
\begin{equation}
	\label{Eq: bdry term operator D}
	\langle \mathbb D_{\mathbf k}\Psi_{\mathbf k}, \Phi_{\mathbf k}\rangle - \langle \Psi_{\mathbf k}, \mathbb D_{\mathbf k} \Phi_{\mathbf k}\rangle = [\overline{\Psi}_{\mathbf k}\Upsilon^1\Phi_{\mathbf k}]_{z=0}=0 \, ,
\end{equation}
where $\overline{\Psi}_{\mathbf k}\doteq \Psi^\dagger_{\mathbf k} \Upsilon^0$ while the $\Upsilon$-matrices are defined in Equation \eqref{eq: upsilon matrices}. This entails that $\mathbb D_{\mathbf k}$ is a symmetric operator and we can look for its self-adjoint extensions by using von Neumann theory of deficiency indices. To this end, we must start by identifying the domain of the adjoint operator, which reads
\begin{equation*}
	\textrm{D}(\mathbb D_{\mathbf k}^\dagger) = \{\Psi_{\mathbf k} \in L^2((0,\infty);\mathbb{C}^{2^{\lfloor\frac{n}{2}\rfloor}}) \; | \; \mathbb D_{\mathbf k}\Psi_{\mathbf k}\in L^2((0,\infty);\mathbb{C}^{2^{\lfloor\frac{n}{2}\rfloor}})\}\, .
\end{equation*}
Subsequently, we need to characterize the deficiency spaces $\mathcal{N}_\pm(\mathbb{D}^\dagger_{\mathbf k})=\ker\left(\mathbb{D}^\dagger_{\mathbf{k}}\pm i\mathbb{I}\right)$ and, to this end, we introduce the following notation to denote certain collections of distinguished elements of $\mathbb C^{\frac{N}{2}}$, with $N=2^{\lfloor\frac{n}{2}\rfloor}$:
\begin{align}
	\label{eq: auxiliarly vectors pads}
	\begin{split}
		&\mathbf{A}= (a_1, \dots, a_{\frac{N}{4}}, 0,\dots,0) \, , \\
		&\mathbf{B}= (b_1, \dots, b_{\frac{N}{4}}, 0,\dots,0) \, , \\
		&\mathbf{C}= (0, \dots, 0, c_1, \dots, c_{\frac{N}{4}}) \, , \\
		&\mathbf{D}= (0, \dots, 0, d_1, \dots, d_{\frac{N}{4}}) \, , 
	\end{split}   
\end{align}
where $\{a_i, \, b_i, \, c_i, \, d_i\}_{i=1,\dots, N/4}$ are arbitrary complex numbers. Consequently, splitting $\Psi =\Psi^{(1)}\oplus \Psi^{(2)}$ as per Section \ref{Sec: gads solution even case}, Equations \eqref{eq: pads SL problem} and \eqref{Eq: lower half solutions dirac eq pads} entail that any element lying in $\mathcal{N}_\pm(\mathbb{D}^\dagger_{\mathbf k})$ can be written as the following linear combinations:
\begin{subequations}
	\label{Eq: general solution pads}
	\begin{equation}\label{Eq: general solution pads-a}
		\Psi^{(1)}_{\pm i,\mathbf{k}}(z) = \sqrt{z}[J_\nu(qz)\mathbf{A}+Y_\nu(qz)\mathbf{B}+J_{\nu-1}(qz)\mathbf{C}+Y_{\nu-1}(qz)\mathbf{D}] \, , 
	\end{equation}
	\begin{gather}
		\Psi^{(2)}_{\pm i,\mathbf{k}}(z) = \mp i\sqrt{z}\{J_\nu(qz)[q\mathbf{C}+ik_j\hat\gamma^j\mathbf{A}] + Y_\nu(qz)[q\mathbf{D}+ik_j\hat\gamma^j\mathbf{B}] + \notag\\
		- J_{\nu-1}(qz)[q\mathbf{A}-ik_j\hat\gamma^j\mathbf{C}] - Y_{\nu-1}(qz)[q\mathbf{B}-ik_j\hat\gamma^j\mathbf{D}] \},\label{Eq: general solution pads-b}
	\end{gather}
\end{subequations}
where we have used implicitly Equations \eqref{Eq: pads solution dirac eq} and \eqref{Eq: lower half solutions dirac eq pads}. In order to deduce whether $\Psi^{(1)}_{\pm i,\mathbf k}(z)$ is square-integrable, it suffices to control the asymptotic expansion as $z\to 0$ or $z\to +\infty$ setting  $\omega= \pm i$. Starting from the former and using \cite[\S 10]{NIST:DLMF} it descends
	\begin{gather}\label{Eq: Asymp1}
		|\Psi^{(1)}_{\pm i,\mathbf{k}}(z)|^2 \stackrel{z\to 0}{\sim} \bigg[\frac{\Im (q)}{2}\bigg]^{-2\nu}\frac{\Gamma^2(\nu-1)}{\pi^2} \Big\{(\nu-1)^2z^{1-2\nu}|\mathbf{B}|^2+  \bigg[\frac{\Im(q)}{2}\bigg]^{2}z^{3-2\nu}|\mathbf{D}|^2\Big\} \ , 
	\end{gather}
where we have reported only the most singular terms. The symbols $\Re(\cdot),\Im(\cdot)$ respectively denote the real and the imaginary part, while $\Gamma(\cdot)$ is the Euler gamma function. As one can infer, the right hand side does not depend on $\omega = \pm i$ and we can distinguish three regimes:
\begin{enumerate}
	\item if $\nu\geq 2$, then $\Psi^{(1)}_{\pm i,\mathbf k}(z)$ is never square-integrable unless $\mathbf B= \mathbf D=0$,
	\item if $1\leq\nu<2$, then $\Psi^{(1)}_{\pm i,\mathbf k}(z)$ is square-integrable only if $\mathbf B=0$,
	\item if $0<\nu<1$, then $\Psi^{(1)}_{\pm i,\mathbf k}(z)$ is always square-integrable.
\end{enumerate}
Accordingly we can study these scenarios separately as $z\to\infty$. More precisely 
\begin{enumerate}
	\item if $\nu\geq 2$, then
	$$\Psi^{(1)}_{\pm i,\mathbf{k}}(z)= \sqrt{z}[J_\nu(qz)\mathbf{A}+J_{\nu-1}(qz)\mathbf{C}] \, . $$
	Since $\sqrt{z}J_\nu(qz)\stackrel{z\to\infty}{\sim}\cos(qz-\frac{i\pi}{2}-\frac{\pi}{4})$ and $q^2=-1-|\mathbf{k}|^2$, using Equation \eqref{eq: auxiliarly vectors pads}, it descends that $\Psi^{(1)}_{\pm i,\mathbf{k}}(z)$ is square-integrable if and only if $\mathbf{A}=\mathbf{C}=0$. 
	\item If $1\leq\nu<2$, then 
	$$\Psi^{(1)}_{\pm i,\mathbf{k}}(z) = \sqrt{z}[J_\nu(qz)\mathbf{A}+J_{\nu-1}(qz)\mathbf{C}+Y_{\nu-1}(qz)\mathbf{D}].$$
	Using the same asymptotic expansion as in the previous point and Equation \eqref{eq: auxiliarly vectors pads}, we can infer that $\mathbf{A}$ must vanish. Accordingly we are left only with a linear combination between the Bessel functions of first and second kind of order $\nu-1$. Provided $\Im(q)>0$, their lone square integrable linear combination is $\sqrt{z}H^{(1)}_{\nu-1}(qz)\stackrel{z\to\infty}{\sim}\exp[iqz-\frac{i\pi}{2}-\frac{\pi}{4}]$, where $H^{(1)}_{\nu-1}(qz)$ denotes the first Hankel function. On the contrary, if $\Im(q)<0$, we can draw the same conclusion provided that we replace the first with the second Hankel function $H^{(2)}_{\nu-1}(qz)$. As a consequence, it turns out that $\mathbf{D}=\pm i\mathbf{C}$ depending on the sign of the imaginary part of $q$. 
	\item If $0<\nu<1$, then no constraint to Equation \eqref{Eq: general solution pads-a} needs to be imposed due to the behaviour close to $z=0$. On the contrary, close to infinity, using the asymptotic expansion of the Bessel functions and observing that Equation \eqref{eq: auxiliarly vectors pads} entails that the vectors $\mathbf{A},\mathbf{B}$ are orthogonal to $\mathbf{C}$ and to $\mathbf{D}$, we can infer that $\Psi^{(1)}_{\pm i,\mathbf{k}}(z)$ is square-integrable provided that $\mathbf{D}=\pm i\mathbf{C}$ and $\mathbf{B}=\pm i\mathbf{A}$, the sign being dependent on that of the imaginary part of $q$ as explained in the previous point.
\end{enumerate} 

\noindent We focus our attention on $\Psi^{(2)}_{\pm i,\mathbf{k}}(z)$ taking into account Equation \eqref{Eq: general solution pads-b} and the constraints on the coefficients $\mathbf{A},\mathbf{B},\mathbf{C}$ and $\mathbf{D}$ deduced from the analysis of $\Psi^{(1)}_{\pm i,\mathbf{k}}(z)$:

\begin{enumerate}
	\item If $\nu\geq 2$, we already know that $\mathbf{A}=\mathbf{B}=\mathbf{C}=\mathbf{D}=0$ and therefore, $\Psi^{(2)}_{\pm i,\mathbf{k}}(z)=0$. In other words $n_\pm(\mathbb{D}_\mathbf k^\dagger)=\dim(\mathcal{N}_\pm(\mathbb{D}^\dagger_{\mathbf k}))=0$, which is tantamount to saying that $\mathbb{D}_\mathbf k$ is essentially self-adjoint.
	\item if $1\leq\nu <2$, the preceding analysis yields $\mathbf{A}=\mathbf{B}=0$ and $\mathbf{C}=\pm i\mathbf{D}$ depending on the sign of $\Im(q)$. Focusing on $\Im(q)>0$, it turns out that 
	$$\Psi^{(2)}_{\pm i,\mathbf{k}}(z) = \mp i\sqrt{z} [qH^{(1)}_\nu(qz) +i H^{(1)}_{\nu-1}(qz)\hat\gamma^jk_j]\mathbf{C}.$$
	Taking into account that, as $z\to 0$, $H^{(1)}_{\nu}(qz)\propto (qz)^{-\nu}$, we need to impose $\mathbf{C}=0$ in order for $\Psi^{(2)}_{\pm i, \mathbf{k}}(z)$ to be square-integrable. Hence $\mathbb{D}_\mathbf k$ is essentially self-adjoint.
	\item If $0<\nu<1$ the preceding analysis yields $\mathbf{A}=\pm i\mathbf{B}$ and $\mathbf{C}=\pm i\mathbf{D}$ depending on the sign of $\Im(q)$. Considering for definiteness $\Im(q)>0$, this entails that 
	\begin{equation*}
		\Psi^{(2)}_\mathbf{k}(z) = \frac{\sqrt{z}}{\omega}\{H^{(1)}_\nu(qz)[q\mathbf{C}+i\hat\gamma^jk_j\mathbf{A}] - H^{(1)}_{\nu-1}(qz)[q\mathbf{A}-i\hat\gamma^jk_j\mathbf{C}] \}  \, ,
	\end{equation*}
	which is square-integrable both at the origin and at infinity. To summarize, in this case, $n_\pm(\mathbb{D}_\mathbf k^\dagger)=\dim(\mathcal{N}_\pm(\mathbb{D}^\dagger_{\mathbf k}))=\frac{N}{2}$ and, hence, on account of von Neumann theory of deficiency indices, the number of self-adjoint extensions is in one-to-one correspondence with the elements of the unitary group $U\left(\frac{N}{2}\right)$.
\end{enumerate}

\noindent To conclude we summarize in Table \ref{tab: expected boundary conditions pads} the results obtained for $\Psi^{(1,2)}_{\pm i, \mathbf k}(z)$:
\begin{table}[htbp]
	\centering
	\begin{tabular}{ccc}
		\toprule
		$\nu$   &  $n_\pm(\mathbb{D}_\mathbf k^\dagger)$ & Expected boundary conditions  \\
		\midrule
		$(0,1)$ & $\frac{N}{2}$ & parametrised by elements of $U\left(\frac{N}{2}\right)$ \\
		$[1,+\infty)$ & $0$ & no conditions \\
		\bottomrule
	\end{tabular}
	\caption{Summary of Von-Neumann theory for the operator $\mathbb D$.}
	\label{tab: expected boundary conditions pads}
\end{table}

We conclude this section by showing how to associate specific boundary conditions to each self adjoint extension of the operator $\mathbb{D}_{\mathbf{k}}$. For definiteness we shall focus on the mass range $\nu\in(0,1)$. As a guiding example we consider first the case $\nu= \frac12$, namely $m=0$. This is a special scenario since all singular terms in Equation \eqref{eq: pads SL problem} vanish and this entails that all solutions of Equation \eqref{eq: pads SL problem} are regular at $z=0$. Therefore one can equivalently write Equation \eqref{Eq: bdry term operator D} as
\begin{equation*}
    [\Phi_{\omega, \mathbf k}^{(1)}(0)+i\hat\gamma^0\Phi_{\omega, \mathbf k}^{(2)}(0)]^\dagger \, [\Psi_{\omega, \mathbf k}^{(1)}(0)+i\hat\gamma^0\Psi_{\omega, \mathbf k}^{(2)}(0)]  = [\Phi_{\omega, \mathbf k}^{(1)}(0)-i\hat\gamma^0\Phi_{\omega, \mathbf k}^{(2)}(0)]^\dagger \, [\Psi_{\omega, \mathbf k}^{(1)}(0)-i\hat\gamma^0\Psi_{\omega, \mathbf k}^{(2)}(0)] \, ,
\end{equation*}
where $\hat{\gamma}^0=i\gamma^0$, see Appendix \ref{Sec: Appendix A}.
If we assume that the boundary values $\Phi_{\omega, \mathbf k}^{(1)}(0)\pm i\hat\gamma^0\Phi_{\omega, \mathbf k}^{(2)}(0)$ are related by a linear transformation, which is indeed the case since the Dirac equation is linear, then it descends that 
\begin{equation*}
    \Phi_{\omega, \mathbf k}^{(1)}(0)+ i\hat\gamma^0\Phi_{\omega, \mathbf k}^{(2)}(0) = U[\Phi_{\omega, \mathbf k}^{(1)}(0) - i\hat\gamma^0\Phi_{\omega, \mathbf k}^{(2)}(0)] \, , \quad U\in U\Big(\frac{N}{2}\Big) \, ,
\end{equation*}
and the same applies for $\Psi_{\omega, \mathbf k}^{(1)}(0)\pm i\hat\gamma^0\Psi_{\omega, \mathbf k}^{(2)}(0)$. Thus, the boundary conditions for $m=0$ are
\begin{equation}	\label{Eq: regular bdry condition pads}
 (\mathbf 1-U)\Psi_{\omega, \mathbf k}^{(1)}(0)+ i(\mathbf 1 + U)\hat\gamma^0\Psi_{\omega, \mathbf k}^{(2)}(0) = 0 \ , 
\end{equation}
where $U\in U\left(\frac{N}{2}\right)$. Let us now assume that $\Phi_{\omega,\mathbf k}(z),\Psi_{\omega,\mathbf k}(z)\in L^2((0,\infty);\mathbb{C}^{2^{\lfloor\frac{n}{2}\rfloor}})$ and that $\Phi_{\omega,\mathbf k}(z)$ also abides by Equation \eqref{Eq: regular bdry condition pads}. Bearing in mind Equation \eqref{Eq: bdry term operator D}, it turns out that 
\begin{gather*}
	\langle \mathbb D_{\mathbf k}\Phi_{\omega,\mathbf k}, \Psi_{\omega,\mathbf k}\rangle - \langle \Phi_{\omega,\mathbf k}, \mathbb D_{\mathbf k} \Psi_{\omega,\mathbf k}\rangle =  \overline{\Phi}_{\omega,\mathbf k}(0) \Upsilon^1 \Psi_{\omega,\mathbf k}(0) = \\
	= \Phi^{(1)\dagger}_{\omega,\mathbf k}(0) \, \hat\gamma^0 \Psi^{(2)}_{\omega,\mathbf k}(0) - \Phi^{(2)\dagger}_{\omega,\mathbf k}(0) \, \hat\gamma^0 \Psi^{(1)}_{\omega,\mathbf k}(0) = \\ 
    = \frac{1}{2i}[\Phi_{\omega, \mathbf k}^{(1)}(0)-i\hat\gamma^0\Phi_{\omega, \mathbf k}^{(2)}(0)]^\dagger \Big\{ U^\dagger[\Psi_{\omega, \mathbf k}^{(1)}(0)+i\hat\gamma^0\Psi_{\omega, \mathbf k}^{(2)}(0)] - [\Psi_{\omega, \mathbf k}^{(1)}(0)-i\hat\gamma^0\Psi_{\omega, \mathbf k}^{(2)}(0)] \Big\}  \, , 
\end{gather*}
where we have exploited that we are working in the regular case and that Equation \eqref{Eq: regular bdry condition pads} holds for $\Phi_{\omega,\mathbf k}$. The right hand side vanishes if 
\begin{equation*}
 (\mathbf 1-U)\Psi_{\omega, \mathbf k}^{(1)}(0)+ i(\mathbf 1 + U)\hat\gamma^0\Psi_{\omega, \mathbf k}^{(2)}(0) = 0 \ . 
\end{equation*}
Hence, we have individuated an extension of $\mathbb{D}_{\mathbf{k}}$ which we denote by $\mathbb{D}^U_{\mathbf{k}}$ such that $\mathbb D^U_{\mathbf{k}}=\mathbb (D^U)^\dagger_{\mathbf{k}}$ and that 
$$\textrm{Dom}(\mathbb{D}^U_{\mathbf{k}})=\{\Psi_{\mathbf k} \in L^2((0,\infty);\mathbb{C}^{2^{\lfloor\frac{n}{2}\rfloor}}) \; |\; (\mathbf 1-U)\Psi_{\omega, \mathbf k}^{(1)}(0)+ i(\mathbf 1 + U)\hat\gamma^0\Psi_{\omega, \mathbf k}^{(2)}(0) = 0\}.$$
Keeping Equation \eqref{Eq: regular bdry condition pads} as an inspiration, we extend our conclusion to the case when $\Psi_{\omega,\mathbf k}(z)$ is singular at $z=0$. To this end, we introduce the $\frac{N}{2}\times\frac{N}{2}$ \textit{rescaling matrix} $R(z) = \text{diag}(z^{|\nu-\frac12|} \mathbf 1, z^{-|\nu-\frac12|} \mathbf 1)$  and we define the \textit{rescaled spinors} as
\begin{equation}
	\label{eq: rescaled spinors}
	_R\Psi_{\omega,\mathbf k}^{(1)}(z) = R(z)\Psi_{\omega,\mathbf k}^{(1)}(z) \, , \quad _R\Psi_{\omega,\mathbf k}^{(2)}(z) = R^{-1}(z)\Psi_{\omega,\mathbf k}^{(2)}(z) \, ,
\end{equation}
where $\Psi_{\omega,\mathbf k}^{(1,2)}(z)$ are as per Equation \eqref{Eq: pads solution dirac eq}. Observe that the rescaled spinor components are regular at $z=0$ since, exploiting Equation \eqref{Eq: general solution pads} for $\nu\in(\frac12, 1)$,
	\begin{align}
		\label{eq: bdry value of rescaled spinors}
		\begin{split}
			\lim_{z\to 0}[ _R\Psi_{\omega,\mathbf k}^{(1)}(z)] &= 
			\begin{pmatrix}
				-\frac{2^\nu\,\Gamma(\nu)}{\pi} \, \mathbf B  \\
				\frac{2^{1-\nu}}{\Gamma(\nu)} \, \mathbf C 
			\end{pmatrix} \, ,\\
			\lim_{z\to 0}[ _R\Psi_{\omega,\mathbf k}^{(2)}(z)] &=
			\begin{pmatrix}
				-\frac{2^\nu\,\Gamma(\nu)}{\pi} \, [q\mathbf{D}+ik_j\hat\gamma^j\mathbf{B}]  \\
				\frac{2^{1-\nu}}{\Gamma(\nu)} \, [q\mathbf{A}-ik_j\hat\gamma^j\mathbf{C}] 
			\end{pmatrix} \ ,
		\end{split}
	\end{align}
	where we have used the asymptotic expansion of the Bessel functions as $z\to 0^+$. The case $\nu\in(0,\frac12)$ is obtained similarly. We remind the reader that, for $\nu\in(0,1)$, an element of Equation \eqref{Eq: general solution pads} lies in the domain of a self-adjoint extension of $\mathbb{D}_{\mathbf{k}}$ once we fix one between the parameters $(\mathbf A, \mathbf B)$ and one between those denoted by $(\mathbf C, \mathbf D)$. Thus, the boundary values of the rescaled spinors encode all data of the solution. 
We show how Equation \eqref{Eq: regular bdry condition pads} can be rewritten in terms of rescaled spinors. For every $\Psi_{\omega, \mathbf k}, \Phi_{\omega,\mathbf k} \in L^2((0,\infty);\mathbb{C}^{2^{\lfloor\frac{n}{2}\rfloor}})$, it holds that
\begin{align*}
	\lim_{z\to0} \{ \overline{\Phi}_{\omega,\mathbf k}(z) \Upsilon^1 \Psi_{\omega,\mathbf k}(z) \}&= \lim_{z\to 0} \{ \Phi^{(1)\dagger}_{\omega,\mathbf k}(z) \, \hat\gamma^0 \Psi^{(2)}_{\omega,\mathbf k}(z) - \Phi^{(2)\dagger}_{\omega,\mathbf k}(z) \, \hat\gamma^0 \Psi^{(1)}_{\omega,\mathbf k}(z) \} = \\
	&= \lim_{z\to 0} \{ _R\Phi^{(1)\dagger}_{\omega,\mathbf k}(z) \, \hat\gamma^0 \, _R\Psi^{(2)}_{\omega,\mathbf k}(z) -  \, _R \Phi^{(2)\dagger}_{\omega,\mathbf k}(z) \, \hat\gamma^0 \, _R\Psi^{(1)}_{\omega,\mathbf k}(z) \Big\} = \\
	&= \, _R\Phi^{(1)\dagger}_{\omega,\mathbf k}(0) \, \hat\gamma^0 \, _R\Psi^{(2)}_{\omega,\mathbf k}(0) -  \, _R \Phi^{(2)\dagger}_{\omega,\mathbf k}(0) \, \hat\gamma^0 \, _R\Psi^{(1)}_{\omega,\mathbf k}(0) \ .
\end{align*}
On account of the rescaling, the evaluation at $z=0$ is well defined and, therefore, we can repeat the same analysis as for $\nu=\frac{1}{2}$ drawing exactly the same conclusions. Hence the most general boundary condition for $\nu\in(0,1)$ which identifies $\mathbb{D}^U_{\mathbf{k}}$, a self-adjoint extension of $\mathbb{D}_{\mathbf{k}}$, reads
\begin{equation}
	\label{eq: singular bdry condition pads}
	(\mathbf 1-U)\,_R\Psi_{\omega, \mathbf k}^{(1)}(0)+ i(\mathbf 1 + U)\hat\gamma^0\, _R\Psi_{\omega, \mathbf k}^{(2)}(0) = 0 \ , 
\end{equation}
where $U\in U(\frac{N}{2})$ while $\hat\gamma^0=i\gamma^0$ is as per Appendix \ref{Sec: Appendix A} and $\Psi^{(1,2)}_{\omega,\mathbf k}\in L^2((0,\infty);\mathbb{C}^{2^{\lfloor\frac{n}{2}\rfloor}})$. 

\begin{remark}\label{remark: MIT-Bag}
	In 1975, while studying hadron models, K. Johnson identified a set of self-adjoint boundary conditions for the Dirac equation on a subset of Minkowski spacetime, see \cite{mit-bag}. These are known as \textit{MIT-bag boundary conditions} and they can be expressed as $n_\mu\Upsilon^\mu \Psi|_{\partial \mathcal M} = \pm \Psi|_{\partial \mathcal M}$, where $\mathcal{M}$ denotes the underlying manifold, $\partial\mathcal M$ its boundary, while $n_\mu$ is an outward pointing, unit vector, orthogonal to $\partial\mathcal M$. In our case the MIT-bag boundary conditions can be written as $\Upsilon^1\, _R\Psi(0) = \pm \,_R\Psi(0)$, where $_R\Psi(0) = _R\Psi^{(1)}(0) \oplus \, _R\Psi^{(2)}(0)$ as per Equation \eqref{eq: rescaled spinors}. \\
    Notice that, starting from Equation \eqref{eq: bdry value of rescaled spinors}, we can recover these distinguished boundary conditions for suitable unitary matrices $U$, though we postpone the discussion to Section \ref{Sec: Ground States}. Henceforth, we refer to Equation \eqref{eq: singular bdry condition pads} as \textbf{generalised MIT-bag boundary condition}.  
\end{remark}
\noindent
We conclude this section by collecting our finding on the admissible self-adjoint boundary conditions in Table \ref{tab: boundary conditions pads}.

\begin{table}[htbp]
    \centering
    \begin{tabular}{cc}
    \toprule
      $\nu$   & Boundary conditions \\
      \midrule 
       $\frac12$  &  $(\mathbf 1-U)\Psi^{(1)}(0)+ i(\mathbf 1 + U)\hat\gamma^0\Psi^{(2)}(0) = 0$\\
       $(0,1)\setminus\{\frac12\}$ & $(\mathbf 1-U)\, _R\Psi^{(1)}(0)+ i(\mathbf 1 + U)\hat\gamma^0\, _R\Psi^{(2)}(0) = 0$\\
       $[1,+\infty)$ & no conditions \\
       \bottomrule
    \end{tabular}
    \caption{Boundary conditions associated to all self-adjoint extensions of $\mathbb D_{\mathbf{k}}$.}
    \label{tab: boundary conditions pads}
\end{table}

\subsection{GAdS$_n$ case}\label{sec: gads boundary conditions}

In this short section we consider the global patch of the $n$-dimensional AdS spacetime, namely GAdS$_n$. Also in this case, we are interested in identifying the admissible boundary conditions which can be supplemented to the Dirac equation \eqref{Eq: Massive Dirac Equation}. Yet, the analysis follows closely the steps already highlighted when considering PAdS$_n$, the Poincar\'e patch, and, therefore, we limit ourselves to stating the main results and highlighting the minor difference with the preceding section. It is worth mentioning that, in \cite{blanco}, a thorough analysis on GAdS$_2$ is available. The starting point consists of writing on GAdS$_n$ the counterpart of the conformal Hamiltonian individuated in Equation \eqref{Eq: conformal dirac hamiltonian pads}. It reads
\begin{equation*} 
	\widetilde H = -\Upsilon^0(\Upsilon^k\nabla_k - m\sec{\rho}) \, , \quad k=1,\dots, n-1,
\end{equation*}
where $\nabla$ denotes the spinor covariant derivative specified for the conformally related manifold $\widetilde{\mathcal M}_n$, see Equation \eqref{eq: spinor covariant derivative} and Section \ref{Sec: Geometry}. Using Equation \eqref{eq: gads anstatz half spinor}, we introduce the counterpart of the operator $\mathbb D_{\mathbf{k}}$ as
\begin{align}
	\label{eq: operator D gads}
	\begin{split}
		&\widetilde H\sum_{\lambda} \underline{f}_{\omega,\lambda}(\rho) \, \underline{\zeta}_\lambda(\rho,\Omega^i) = \sum_{\lambda} [\widetilde{\mathbb D}\underline{f}_{\omega,\lambda}(\rho)] \, \underline{\zeta}_\lambda(\rho,\Omega^i) \, , \\
		&\widetilde{\mathbb D} \doteq 
		\begin{pmatrix}
			0 & 0 & -\frac{d}{d\rho}+m\sec{\rho} & 0 \\
			0 & 0 & 0 &   \frac{d}{d\rho}+m\sec{\rho}  \\
			\frac{d}{d\rho}+m\sec{\rho} & 0 & 0 & 0 \\
			0 &  -\frac{d}{d\rho}+m\sec{\rho} & 0  & 0
		\end{pmatrix} \, ,
	\end{split}
\end{align}
where $\underline{f}_{\omega,\lambda}(\rho) \, \underline{\zeta}_\lambda(\rho,\Omega^i)$ denotes a vector whose components are split in two halves. The first takes the form $e^{\mp i\lambda\rho}f^{(1,\pm)}_{\omega,\lambda}(\rho) \, \zeta_\lambda(\rho,\Omega^i)$ while the second one $e^{\mp i\lambda\rho}f^{(2,\pm)}_{\omega,\lambda}(\rho) \, \zeta_\lambda(\rho,\Omega^i)$. Notice how Equation \eqref{eq: gads anstatz half spinor} entails that $\widetilde{\mathbb D}$ is a $4\times4$ matrix regardless of the dimension of GAdS$_n$, with $n>3$. Once more, $\widetilde{\mathbb D}$ as per Equation \eqref{eq: operator D gads} can be regarded as a higher-dimensional generalisation of the one discussed for $n=2$ in \cite{blanco}, which in turn also applies to the scenario with $n=3$.  The induced inner product starting from Equations \eqref{Eq: GAdSn inner product} and \eqref{Eq: ESU inner product} reads
\begin{equation*}
	\langle f_{\omega, \lambda}, g_{\omega, \lambda}\rangle = \sum_\lambda \int_{-\frac{\pi}{2}}^{+\frac{\pi}{2}} d\rho \, f_{\omega, \lambda}^\dagger(\rho)g_{\omega, \lambda}(\rho) \, ,
\end{equation*}
where $f_{\omega, \lambda}, g_{\omega, \lambda} \in L^2((0,\frac{\pi}{2});\mathbb{C}^{2^{\lfloor\frac{n}{2}\rfloor}})$ have respectively components $f^{(1,\pm)}, f^{(2,\pm)}$ and $g^{(1,\pm)}, g^{(2,\pm)}$. As one could infer, $\widetilde{\mathbb D}$ is symmetric with respect to this inner product only if it vanishes
\begin{equation}
	\label{eq: bdry term operator D gads}
	\langle \widetilde{\mathbb D} f_{\omega, \lambda}, g_{\omega, \lambda}\rangle - \langle  f_{\omega, \lambda},\widetilde{\mathbb D} g_{\omega, \lambda}\rangle \propto \sum_\lambda \, [\overline{f}_{\omega, \lambda}\Upsilon^1 g_{\omega, \lambda}]^{\frac{\pi}{2}}_{0} \, ,
\end{equation}
where $\overline{f} = f^\dagger \Upsilon^0$. It suffices to consider $C^\infty_0(0,\frac{\pi}{2})$ to ensure that this holds true and hence to conclude that $\widetilde{\mathbb D}$ is a symmetric operator on $L^2(0,\frac{\pi}{2})$. In order to characterize the self-adjoint extensions of $\widetilde{\mathbb D}$, one should proceed exactly as in the previous section. Since mutatis mutandis the procedure is identical, we list only the final result:
\begin{itemize}
	\item If $\nu\doteq m  + \frac12\in (0,1)$ the deficiency indices associated to $\widetilde{\mathbb D}$ and to its adjoint $\widetilde{\mathbb D}^\dagger$ are $n_\pm(\widetilde{\mathbb D}^\dagger) = 4$.
	\item If $\nu\geq 1$ $\widetilde{\mathbb D}$ is essentially self-adjoint, thus $n_\pm(\widetilde{\mathbb D}^\dagger) = 0$.
\end{itemize}
To conclude, we exploit once more the analogy with Section \ref{sec: pads boundary conditions} and \cite{blanco} in order to write the sought boundary conditions on GAdS$_n$:
\begin{itemize}
	\item For $\nu\in(0,1)$ we introduce the rescaling factor $R(\rho) = \sigma(\rho)^{|\nu-\frac12|}$, where $\sigma(\rho)$ is as per Equation \eqref{gads solutions list}. The boundary conditions are then given by
	\begin{equation}
		\label{eq: singular bdry condition gads}
		(\mathbf 1 - U) 
		\begin{pmatrix}
			_R f^{(2)}(\frac{\pi}{2}) \\
			_R f^{(2)}(0)
		\end{pmatrix}
		= i(\mathbf 1 + U) 
		\begin{pmatrix}
			_R f^{(1)}(\frac{\pi}{2}) \\
			_R f^{(1)}(0)
		\end{pmatrix} \, ,
	\end{equation}
	where $U\in U(4)$, while
	\begin{equation}
		_R f^{(1)}(\rho) = R(\rho)f^{(1)}(\rho) \, , \quad _R f^{(2)}(\rho) = R^{-1}(\rho)f^{(2)}(\rho) \, .
	\end{equation}
	
	\item For $\nu\geq 1$, $\mathbb D$ is essentially self-adjoint. Thus, we do not need to impose any boundary condition on the square-integrable spinors.
\end{itemize}
Also in this scenario, a suitable choice of $U$ is tantamount to selecting boundary conditions as in the MIT-bag model \cite{mit-bag}.

\section{Green's operator for the Dirac equation}\label{sec: green operator dirac equation}

In this section we derive an expression for the advanced $G^-$ and retarded $G^+$ fundamental solutions for the Dirac equation on PAdS$_n$ for each of the boundary conditions which have been individuated in Section \ref{Sec: Solutions pads dirac eq even}. We focus on this scenario since on GAdS$_n$ the computations are analogous, see Sections \ref{sec: pads boundary conditions} and \ref{sec: gads boundary conditions}. We recall that a fundamental solution is a map $\mathcal{G}^\pm:C^\infty_0(\mathbb{R}^+\times\mathbb{R}^{n-1};\mathbb{C}^{2^{\lfloor\frac{n}{2}\rfloor}})\to C^\infty(\mathbb{R}^+\times\mathbb{R}^{n-1};\mathbb{C}^{2^{\lfloor\frac{n}{2}\rfloor}})$ such that
\begin{equation}\label{Eq: Defining Green}
	P\circ \mathcal{G}^\pm = \mathcal{G}^\pm \circ P|_{C^\infty_0} = \mathrm{Id}_{C^\infty_0} \, ,
\end{equation}
where $P$ is the Dirac operator as per Equation \eqref{Eq: Massive Dirac in PAdS}. Equivalently, we look for a bi-distribution $G^\pm\in\mathcal{D}^\prime((\mathbb{R}^+\times\mathbb{R}^{n-1})\times(\mathbb{R}^+\times\mathbb{R}^{n-1});\mathbb{C}^{2^{\lfloor\frac{n}{2}\rfloor}}\otimes \mathbb{C}^{2^{\lfloor\frac{n}{2}\rfloor}})$ such that, working at the level of integral kernels,
\begin{equation}
	\label{eq: def of green operator with kernels}
	P_x G^{\pm}(x,x')=  G^\pm(x,x')\, \overline{P'_{x'}} = \delta(x-x')\mathsf{1}_{\mathbb{C}^{2^{\lfloor\frac{n}{2}\rfloor}}}\, ,       
\end{equation}
where $\delta(x-y)$ is the Dirac $\delta$-distribution on $\mathbb{R}^+\times\mathbb{R}^{n-1}$, $P_x$ is the Dirac operator as per Equation \eqref{Eq: Massive Dirac in PAdS} acting on the $x$-entry, while $\overline{P'_{x'}}$ is the anti-Dirac operator as per Equation \eqref{eq: anti-dirac equation} acting on the $x'$-entry. Once more, it is convenient to consider a conformally equivalent problem on  $\mathbb H^n$, see Section \ref{Sec: Solutions pads dirac eq even}. To establish the correspondence, we recall that any $\psi:\mathbb{R}^+\times\mathbb{R}^{n-1}\to \mathbb{C}^{2^{\lfloor\frac{n}{2}\rfloor}}$ is conformally related to $\Psi :=z^\frac{1-n}{2}\psi:\mathbb{H}^n\to\mathbb{C}^{2^{\lfloor\frac{n}{2}\rfloor}}$ such that $P\psi = z^{\frac{n+1}{2}}\widetilde P\Psi$ where $P$ is the Dirac operator on PAdS$_n$ while $\widetilde{P}$ is defined in Equation \eqref{Eq: conformal + change basis pads dirac eq}. Equation \eqref{Eq: Defining Green} entails that, denoting by $\widetilde{G}^\pm$ the retarded and advanced fundamental solutions associated to $\widetilde{P}$, these are related to $G^\pm$ via the identity
\begin{equation*}
\widetilde{G}^\pm=z^{\frac{n-1}{2}}\circ G^\pm\circ z^{-\frac{n+1}{2}}.
\end{equation*}
Hence we can focus our attention to the construction of the retarded fundamental solutions directly on $\mathring{\mathbb{H}}^n$, so to avoid considering the test-functions whose support intersects the boundary at $z=0$. The most notable advantage is that the identity 
\begin{equation}
\label{eq: dirac - anti dirac operator}
	S\doteq P\circ P' = \Big(\Upsilon^\nu\partial_\nu+\frac{m}{z}\mathsf{1}\Big)\circ\Big(\Upsilon^\mu\partial_\mu-\frac{m}{z}\mathsf{1}\Big) =  \eta^{\mu\nu}\partial_\mu\partial_\nu\mathsf{1} + \Upsilon^1 \frac{m}{z^2}- \frac{m^2}{z^2}\mathsf{1} \, ,
\end{equation}
entails that we construct the advanced and retarded fundamental solution $\mathcal{K}^\pm$ for $S$ since 
\begin{equation}\label{eq: induced green operator}
\widetilde{G}^\pm=\Big(\Upsilon^\nu\partial_\nu+\frac{m}{z}\mathsf{1}\Big)\mathcal{K}^\pm.
\end{equation}
In turn, since $\Upsilon^1 = \text{diag}(-\mathbf 1, \mathbf 1, \mathbf 1, -\mathbf 1)$ as per Equation \eqref{eq: upsilon matrices}, we can infer that 
\begin{equation}
	\label{Eq: Kpm}
	\mathcal{K}^\pm= \begin{pmatrix}
		\widetilde K^\pm_{\nu}\mathbf 1_{\mathbb{C}^{2^{\lfloor\frac{n}{2}\rfloor-1}}}  & 0 &0 &0 \\
		0 & \widetilde K^\pm_{\nu-1}\mathbf 1_{\mathbb{C}^{2^{\lfloor\frac{n}{2}\rfloor-1}}} & 0 & 0 \\
		0 & 0 & \widetilde K^\pm_{\nu-1}\mathbf{1}_{\mathbb{C}^{2^{\lfloor\frac{n}{2}\rfloor-1}}} & 0 \\
		0 & 0 & 0 & \widetilde K^\pm_{\nu} \mathbf{1}_{\mathbb{C}^{2^{\lfloor\frac{n}{2}\rfloor-1}}}
	\end{pmatrix} \, ,
\end{equation}
where $\widetilde K^\pm_\nu$ are the advanced and retarded fundamental solution of the Klein-Gordon equation on $\mathring{\mathbb{H}}^n$ with mass $\nu=m+\frac{1}{2}$. We denote by $\mathcal K (x,x')$ the integral kernel of the advanced-minus-retarded Green's operator for $S$, see \cite{dappiaggi1}, and by $\mathcal{G} (x,x')$ its counterpart for $P$ as per Equation \eqref{eq: induced green operator}. The former can be determined once we fix any admissible boundary condition that is Equation \eqref{Eq: Kpm} ought to satisfy Equation \eqref{eq: singular bdry condition pads}. In order to translate this statement in a condition that $\widetilde K_\nu(x,x')$ has to abide by, we proceed as follows. Taking once more \cite{dappiaggi1} as an inspiration and using the notation introduced in Equation \eqref{eq: pads fourier transform}, we write
\begin{equation}
	\label{eq: fourier expansion adv-ret green operator}
	\mathcal G(x,x') =  \int_{\mathbb R^{n-2}}  \frac{d\mathbf k} {(2\pi)^{\frac{n-2}{2}}} \, e^{-i\mathbf k \cdot (\mathbf x- \mathbf x')} \, \int_{-\infty}^{+\infty} \frac{d\omega}{\sqrt{2\pi}} \, e^{i\omega (t-t')} \mathbf G_{\omega, \mathbf k}(z,z') \, ,
\end{equation}
where $\mathcal G^-=\Theta(t^\prime-t)\mathcal G$, $\mathcal{G}^+=-\Theta(t^\prime-t)\mathcal G$ and 
\begin{subequations}
	\begin{equation}
			 P_x \mathcal G(x,x') = 0 \, , \label{eq: dirac eq for adv-ret green operator} 
	\end{equation}
		\begin{equation}
	\mathcal G(x,x') \, \overline{P'_{x'}} = 0 \, , \label{eq: anti-dirac eq for adv-ret green operator}
	\end{equation}
	\begin{equation}
	\mathcal G(x,x')|_{t=t'} = \Upsilon^0 \delta(z-z')\delta(\mathbf x-\mathbf x')  \, , \label{eq: equal time adv-ret green operator}
	\end{equation}
\end{subequations}
in addition to a boundary condition as per Table \ref{tab: boundary conditions pads}. Replacing Equation \eqref{eq: fourier expansion adv-ret green operator} in Equation \eqref{eq: equal time adv-ret green operator} yields
\begin{equation}
	\label{eq: eigenfunction expansion adv-ret green operator}
	\int_{-\infty}^{+\infty} \frac{d\omega}{\sqrt{2\pi}} \, \mathbf G_{\omega, \mathbf k}(z,z') = \Upsilon^0 \, \delta(z-z') \, .
\end{equation}
As one can infer from Equations \eqref{eq: fourier expansion adv-ret green operator}, \eqref{eq: dirac eq for adv-ret green operator} and \eqref{eq: anti-dirac eq for adv-ret green operator}, the columns of $\mathbf G_{\omega, \mathbf k}(z,z')$, regarded as functions of $z$, are eigenfunctions of $\mathbb D_{\mathbf{k}}$ associated to the eigenvalue $\omega$, while its rows, regarded as functions of $z'$, are conjugated eigenfunctions associated to the same eigenvalue. In other words, Equation \eqref{eq: eigenfunction expansion adv-ret green operator} establishes a spectral representation of the $\delta$-distribution in terms of eigenfunctions of $\mathbb D_{\mathbf{k}}$ subject to boundary conditions as per Table \ref{tab: boundary conditions pads}. 

\paragraph{Constructing $\mathbf G_{\omega, \mathbf k}(z,z^\prime)$ --} Equation \eqref{eq: eigenfunction expansion adv-ret green operator} entails that, barring a factor $\Upsilon^0$, $\mathbf G_{\omega, \mathbf k}(z,z')$ can be constructed out of $\mathbb{D}_{\mathbf{k}}$ as per Equation \eqref{eq: operator D pads} solving
\begin{equation}
	\label{eq: green operator equation for operator D}
	(\mathbb D_{\mathbf{k}} + \omega) G_{\mathbb D_{\mathbf{k}}}(z,z';\omega)= \delta(z-z') \, ,
\end{equation}
together with one among the boundary conditions as per Table \ref{tab: boundary conditions pads}. We recall that $\mathbb D_{\mathbf{k}}$ in Equation \eqref{eq: green operator equation for operator D} is acting on $z$ and we write $G_{{\mathbb D}_{\mathbf{k}}}(z,z';\omega)$ as
\begin{equation}\label{eq: piecewise green function}
G_{{\mathbb D}_{\mathbf{k}}}(z,z';\omega) = 
    \begin{cases}
        \mathscr F_\omega(z)\mathcal C_<(z') \ , \ z<z' \ , \\
        \mathscr F_\omega(z)\mathcal C_>(z') \ , \ z>z' \ ,
    \end{cases}
\end{equation}
where, setting $N=2^{\lfloor\frac{n}{2}\rfloor}$, $\mathscr F_\omega(z)$ is the $N\times N$ matrix whose columns are the solutions listed in Equations \eqref{Eq: pads solution dirac eq}, \eqref{Eq: lower half spinor pads}. On the contrary $\mathcal C_{\gtrless}(z')$ are $N\times N$ matrices to be determined from Equation \eqref{eq: green operator equation for operator D} and from the boundary conditions listed in Table \ref{tab: boundary conditions pads}. In the following we investigate more in detail the conditions that they have to fulfill, splitting the analysis between two scenarios depending on $\nu$ and, thus, on the mass of the field. We highlight that, in this section, we do not determine explicitly their expression, rather we establish the algorithm to determine then. In Section \ref{Sec: Ground States} we will apply it to two concrete cases. 

\paragraph{Case $\mathbf{\nu\in(0,1)}$ --} For $0<z<z'<+\infty$ and denoting by $U$ a matrix which identifies an admissible boundary condition, $G_{\mathbb D_{\mathbf{k}}}(z,z';\omega)$ has to fulfill
\begin{equation*}
	(\mathbf 1 - U)\, _RG_{\mathbb D_{\mathbf{k}}}^{(1)}(0,z';\omega) + i(\mathbf 1 + U)\hat\gamma^0\, _RG_{\mathbb D_{\mathbf{k}}}^{(2)}(0,z';\omega) = 0 \ , 
\end{equation*}
where $_RG_{\mathbb D_{\mathbf{k}}}^{(1,2)}(z,z';\omega)$ denote the rescaled upper/lower half the matrix $G_{\mathbb D_{\mathbf{k}}}(z,z';\omega)$ as per Equation \eqref{eq: rescaled spinors}. Equation \eqref{eq: piecewise green function} yields
\begin{equation*}
    	\big\{(\mathbf 1 - U)\, [_R\mathscr F_\omega (0)]^{(1)} + i(\mathbf 1 +U)\hat\gamma^0\, [_R\mathscr F_\omega (0)]^{(2)}\big\}  \,\mathcal C_<(z') = 0 \ .
\end{equation*}
For $0<z'<z<+\infty$ no boundary condition must be imposed and, following \cite{green-function-bvp}, we can infer that the columns of the matrix $\mathscr F_\omega(z)C_>(z')$ must be eigenfunctions of $\mathbb D_{\mathbf{k}}$ that are square-integrable when $\text{Im}(\omega) \neq 0$. The analysis carried out on $\Psi_{\pm i}(z)$, see Section \ref{Sec: Solutions pads dirac eq even}, suggests that
\begin{equation*}
 [\mathcal C_>(z')]^{(2)} = \pm i[\mathcal C_>(z')]^{(1)} \, ,
\end{equation*}
where the sign is the same as that of $\text{Im}(\omega)$. At last, we impose a matching condition at $z=z'$ as follows. We consider Equation \eqref{eq: green operator equation for operator D} and we integrate over $z\in(z'-\varepsilon, z'+\varepsilon)$ for $\varepsilon>0$. Given $\mathbb D_{\mathbf{k}}$ as per Equation \eqref{eq: operator D pads}, we take the limit $\varepsilon\to0$ so that it holds
\begin{equation*}
	\lim_{\varepsilon\to0}  \big[ G(z'+\varepsilon, z';\omega)|_{z>z'} - G(z'-\varepsilon, z';\omega)|_{z<z'}\big] = -i\Upsilon^0\Upsilon^1 \, .
\end{equation*}
This identity translates to 
\begin{equation*}
	\mathcal C_>(z') -\mathcal C_<(z') = -i\,\mathscr F^{-1}_\omega(z') \,\Upsilon^0\Upsilon^1 \, .
\end{equation*}
Gathering all the previous constraints above, one can conclude that the matrices $\mathcal C_\gtrless(z')$ abide by
\begin{equation} \label{eq: general system for green function}
    \begin{cases}
    [\mathcal C_>(z')]^{(2)} = \pm i[\mathcal C_>(z')]^{(1)} \ , \\
        \{(\mathbf 1 -U)\, [_R\mathscr F_\omega (0)]^{(1)} + i(\mathbf 1 + U)\hat\gamma^0\, [_R\mathscr F_\omega (0)]^{(2)}\}  \,[\mathcal C_>(z')+i\,\mathscr F_\omega^{-1}(z')\Upsilon^0\Upsilon^1] = 0 \ , \\
        \mathcal C_<(z') = \mathcal C_>(z') +i\, \mathscr F^{-1}_\omega(z') \,\Upsilon^0\Upsilon^1  \, .
    \end{cases}
\end{equation}

\paragraph{Case $\mathbf{\nu\in[1,+\infty)}$ --} This case is simpler since we do not have to impose boundary conditions at $z=0$. Hence, we need to consider only square-integrable spinors, namely $\mathbf B  = \mathbf D = 0$ as per Section \ref{sec: pads boundary conditions}, or, equivalently, $[\mathcal C_<(z')]^{(2)} = 0$. Apart from this difference, the analysis is the same as in the previous case and we do not report it again. 

\paragraph{Resolution of the Dirac Delta --} Following Equation \eqref{eq: eigenfunction expansion adv-ret green operator}, one must read $G_{{\mathbb D}_{\mathbf{k}}}(z,z',\omega)$ as a function of the spectral parameter $\omega$. Yet Equation \eqref{eq: piecewise green function} entails that $G_{{\mathbb D}_{\mathbf{k}}}$ depends on $\mathscr F_\omega(z)$ and on $\mathcal C_{\gtrless}(z')$. Combining Equations \eqref{eq: general system for green function} and \eqref{eq: pads SL problem}, it descends that, rather than on $\omega$, it is more convenient to work with $q$ where $q^2=\omega^2-|\mathbf{k}|^2$. Repeating the same analysis outlined in \cite{dappiaggi1} based in turn on the theory of Sturm-Liouville equations, see also  \cite{green-function-bvp}, it descends that 
\begin{align}\label{eq: definition of jump across branch cut}
    \delta(z-z') = -\frac{1}{2\pi i}\lim_{\varepsilon\to 0^+}\int_{\sigma(\mathbb D_{\mathbf{k}})}[G_{{\mathbb D}_{\mathbf{k}}}(z,z',\omega+i\varepsilon) - G_{{\mathbb D}_{\mathbf{k}}}^\dagger(z',z,\omega-i\varepsilon)]d\omega \, ,
\end{align}
where $G_{{\mathbb D}_{\mathbf{k}}}(z,z',\omega\pm i\varepsilon)$ denote respectively the Green's functions of the operator ${\mathbb D}_{\mathbf{k}}-\omega\mp i\varepsilon$ while $\sigma(\mathbb D_{\mathbf{k}})$ denotes the spectrum of $\mathbb D_{\mathbf{k}}$.     Following \cite{green-function-bvp}, in order to give a representation of Equation \eqref{eq: definition of jump across branch cut}, we look for $\{u_\omega(z)\}_{\omega\in\sigma(\mathbb D_{\mathbf{k}})}$, a complete set of generalized eigenfunctions for $\mathbb D_{\mathbf{k}}$, which satisfy the underlying chosen boundary condition as per Table \ref{tab: boundary conditions pads} as well as the identity
     \begin{equation*}
        \int_{\sigma(\mathbb D_{\mathbf{k}})} u_\omega(z)  u_\omega^\dagger(z')d\omega = \delta(z-z') \, .
    \end{equation*}
A direct comparison of this identity with Equation \eqref{eq: eigenfunction expansion adv-ret green operator} allows to determine $G_{{\mathbb D}_{\mathbf{k}}}$.

\section{Propagators for the Dirac Field}\label{Sec: Ground States}
In this Section we apply the algorithm developed in Section \ref{sec: green operator dirac equation} to a massive Dirac field and, for definiteness, we consider only PAdS$_4$ although the following analysis can be adapted to any spacetime dimension $n\geq 2$. In particular, we focus on two boundary conditions as per Table \ref{tab: boundary conditions pads}, one being the \textit{MIT-Bag boundary condition}, namely $\Upsilon^1\,_R\Psi(0) = \pm\,_R\Psi(0)$ as per Remark \ref{remark: MIT-Bag} and one being a generalized counterpart. In what follows we assume for definiteness $m\in(0,\frac{1}{2})$, \textit{i.e.}, $\nu\in(\frac12,1)$.

\paragraph{Case 1: Upper-half generalized MIT-Bag condition}\label{sec: adv-ret green with pole}
We start our investigation by setting $_R\Psi^{(1)}(0) = 0$ which, according to Table \ref{tab: boundary conditions pads}, corresponds to choosing $U = -\mathbf 1$. We refer to it as {\em Upper-half generalized MIT-Bag condition}. Recalling that, in this setting, $\mathbf k = (k_1, k_2)$ and using Equations \eqref{eq: pads SL problem} and \eqref{eq: pads system dirac eq}, we can infer that
\begin{equation}\label{eq: fundamental matrix}
\mathscr F_\omega(z) = \frac{\sqrt{z}}{\omega} \left( \begin{smallmatrix}
		\omega J_\nu(qz) & 0 & \omega Y_\nu(qz) & 0 \\
		0 &  \omega J_{\nu-1}(qz) & 0 &  \omega Y_{\nu-1}(qz) \\
		-qJ_{\nu-1}(qz) &  (ik_1+k_2)J_{\nu-1}(qz) & -qY_{\nu-1}(qz) &   (ik_1+k_2)Y_{\nu-1}(qz) \\
		  (ik_1-k_2)J_\nu(qz) & -qJ_{\nu}(qz) &   (ik_1-k_2)Y_\nu(qz) & -qY_{\nu} (qz)
	\end{smallmatrix} \right) \, .
\end{equation}
\begin{remark}
 The columns of $\mathscr F_\omega(z)$ are, by construction, the four linear independent solutions of the Dirac equation on $PAdS_4$. Comparing with Equation \eqref{eq: bdry value of rescaled spinors}, one can infer that the first and the last columns are those implementing the chosen boundary condition. As a matter of fact the first one satisfies $\Upsilon^1\,_R\Psi(0) = \,_R\Psi(0)$ while the last one $\Upsilon^1\,_R\Psi(0) = -\,_R\Psi(0)$. Hence, $_R\Psi^{(1)}(0) = 0$ and we stress that this is a boundary condition \textbf{distinguished} from those discussed in \cite{mit-bag}.
\end{remark}
\noindent
The next step consists of focusing on Equation \eqref{eq: general system for green function} which entails that the matrices $\mathcal C_{\gtrless}(z')$ solve
\begin{equation}\label{eq: specific system for green function}
\begin{cases}
    [\mathcal C_>(z')]^{(2)} = \pm i[\mathcal C_>(z')]^{(1)} \ , \\
    [_R\mathscr F_\omega (0)]^{(1)} \,[\mathcal C_>(z')+i\,\mathscr F_\omega^{-1}(z')\Upsilon^0\Upsilon^1] = 0 \ , \\
        \mathcal C_<(z') = \mathcal C_>(z') +i\, \mathscr F^{-1}_\omega(z') \,\Upsilon^0\Upsilon^1  \, ,
\end{cases}
\end{equation}
where the sign in the first identity is determined by that of $\text{Im}(\omega)$, as per Section \ref{sec: green operator dirac equation}. A direct computation yields
\begin{gather*}
    C_>^{(\pm)}(z') = \frac{\pi\sqrt{z'}}{2} \left( \begin{smallmatrix}
       \mp i\omega J_\nu(qz) & 0 & \pm iq J_{\nu-1}(qz) & \pm (ik_2-k_1)J_\nu(qz) \\
       0 & -\omega Y_{\nu-1}(qz) & (ik_1-k_2)Y_{\nu-1}(qz)  & qY_\nu(qz) \\
       \omega J_\nu(qz) & 0 & -q J_{\nu-1}(qz) & -(ik_1+k_2)J_\nu(qz) \\
        0 & \mp i\omega Y_{\nu-1}(qz) & \mp (ik_2+k_1)Y_{\nu-1}(qz)  & \pm iqY_\nu(qz)
    \end{smallmatrix} \right)\ ,\\ \\
    C_<^{(\pm)}(z') = \frac{\pi\sqrt{z'}}{2}\left(\begin{smallmatrix}
       \mp i\omega H^{(\pm)}_\nu(qz) & 0 & \pm iq H^{(\pm)}_{\nu-1}(qz)  & \pm (ik_2-k_1)H^{(\pm)}_\nu(qz) \\
       0 & 0 & 0  & 0 \\
       0 & 0 & 0 & 0 \\
        0 & -\omega H^{(\pm)}_{\nu-1}(qz) & (ik_1-k_2)H^{(\pm)}_{\nu-1}(qz)  & qH^{(\pm)}_{\nu}(qz)
    \end{smallmatrix} \right)\ ,
\end{gather*}
where, with a slight abuse of notation, we employ the symbols $H_\alpha^{(\pm)}(x) = J_\alpha(x) \pm iY_\alpha(x)$ to denote the Hankel functions respectively of first and second kind. Using Equation \eqref{eq: piecewise green function} one can construct  $G_{\mathbb{D}_\mathbf k}(z,z',\omega)$ though we avoid reporting its explicit form since it is quite heavy to read and not particularly informative. On the contrary we focus on Equation \eqref{eq: definition of jump across branch cut} which, removing the $\varepsilon$-dependence from the integrand, translates to 
\begin{equation}\label{eq: spectral density}
\begin{gathered}
    \rho_{{\mathbb D}_{\mathbf{k}}}(\omega) \doteq -\frac{1}{2\pi i}\left(G_{{\mathbb D}_{\mathbf{k}}}(z,z',\omega) - G_{{\mathbb D}_{\mathbf{k}}}^\dagger(z',z,\omega)\right) = \\
    \frac{\sqrt{zz'}}{2}\Theta(\omega^2-|\mathbf{k}|^2)
    \left( \begin{smallmatrix}
	  \omega J^\nu_{\nu}(z,z') & 0 & -q J^\nu_{\nu-1}(z,z') & -\widetilde k J^\nu_{\nu}(z,z') \\
      0 & \omega Y_{\nu-1}^{\nu-1}(z,z') &  \widetilde k^\star \, Y_{\nu-1}^{\nu-1}(z,z') & -q Y^{\nu-1}_{\nu}(z,z') \\
      -q J^{\nu-1}_{\nu}(z,z') & -i\widetilde k \,Y_{\nu-1}^{\nu-1}(z,z') & \frac{1}{\omega}(J_{\nu-1}\star Y_{\nu-1})(z,z')& \frac{q}{\omega}\widetilde k [J^{\nu-1}_{\nu}(z,z') - Y^{\nu-1}_{\nu}(z,z')] \\
      -\widetilde k^\star J_{\nu}^{\nu}(z,z') & -q Y^{\nu}_{\nu-1}(z,z') & \frac{q}{\omega}\widetilde k^\star [J^{\nu}_{\nu-1}(z,z') - Y^{\nu}_{\nu-1}(z,z')]& \frac{1}{\omega}(Y_\nu \star J_\nu)(z,z')
	\end{smallmatrix} \right )  \ ,
\end{gathered}
\end{equation}
where we have employed the following compact notation
\begin{gather*}
    J^\alpha_{\beta}(z,z') = J_\alpha(qz)J_\beta(qz') \, , \quad   Y^\alpha_{\beta} = Y_\alpha(qz)Y_\beta(qz')  \, ,\\
    (J_\alpha\star Y_\beta)(z,z') = q^2J^\alpha_{\alpha}(z,z') + |\mathbf k|^2Y^\beta_{\beta}(z,z') \, , \\
    \widetilde k = ik_1 + k_2 \, , \quad \widetilde{k}^\star = -ik_1 + k_2 \, .
\end{gather*}
Notice that $|\widetilde{k}| = |\mathbf k|$. One can verify by direct inspection that 
\begin{equation*}
    \rho_{\mathbb D_\mathbf k}(\omega) = u_1(z)u_1(z')^\dagger + u_2(z)u_2(z')^\dagger \ ,
\end{equation*}
where $\{u_1,u_2\}$ is a set of complete eigenfunctions that fulfill the upper-half MIT-Bag boundary conditions:
\begin{equation*}
    u_1(z) = \frac{1}{\sqrt{2\omega}}\begin{bmatrix}
       \omega J_\nu(qz) \\
       0 \\
       -qJ_{\nu-1}(qz) \\
       (ik_1-k_2)J_\nu(qz)
    \end{bmatrix}, \quad
    u_2(z) = \frac{1}{\sqrt{2\omega}}\begin{bmatrix}
       0 \\
       \omega Y_{\nu-1}(qz) \\
       (ik_1+k_2)Y_{\nu-1}(qz) \\
       -qY_{\nu}(qz)        
    \end{bmatrix} \ .
\end{equation*}
As one can infer both from this expression and from Equation \eqref{eq: spectral density}, there is a simple pole at $\omega=0$. A direct computation yields that the residue associated to it is
\begin{equation}\label{eq: residue}
\begin{gathered}
    \rho^{dis}_{\mathbb D_{\mathbf{k}}} (0) = |\mathbf k|^2\sin(\pi\nu)\frac{2\sqrt{zz'}}{\pi}
   \left( \begin{smallmatrix}
	   0 & 0 & 0 & 0 \\
      0 & 0 & 0 & 0  \\
      0 & 0  & K_{\nu-1}(|\mathbf k|z)K_{\nu-1}(|\mathbf k|z') & \frac{ik_1 + k_2}{|\mathbf k|}K_{\nu-1}(|\mathbf k|z) K_{\nu}(|\mathbf k|z') \\
      0 & 0 & \frac{-ik_1+k_2}{|\mathbf k|}K_\nu(|\mathbf k|z) K_{\nu-1}(|\mathbf k|z') & K_\nu(|\mathbf k|z) K_\nu(|\mathbf k|z')
	\end{smallmatrix} \right ) \ ,
    \end{gathered}
\end{equation}
where $K_\alpha(x)$ is the modified Bessel function of second kind. This is tantamount to considering the square-integrable eigenfunction of $\mathbb D_{\mathbf{k}}$, associated to the eigenvalue $\omega = 0$, which fulfills the upper-half MIT-Bag boundary condition, namely
\begin{equation*}
\rho^{dis}_{\mathbb D_{\mathbf{k}}}(0) = v(z)v(z)^\dagger \, , \quad 
    v(z) = |\mathbf k|\sqrt{\frac{2\sin(\pi\nu)z}{\pi}}\begin{bmatrix}
       0 \\
       0 \\
       K_{\nu-1}(|\mathbf k|z) \\
      -\frac{ik_1-k_2}{|\mathbf k|}K_\nu(|\mathbf k|z)
    \end{bmatrix} \, .
\end{equation*}
Putting together all these data and taking into account Remark \ref{remark: zero mode solution}, the spectral decomposition of the $\delta$-distribution reads
\begin{equation}
    \delta(z-z') = \rho^{dis}_{\mathbb D_{\mathbf{k}}}(0) + \int_\mathbb{R} \rho_{\mathbb D_{\mathbf{k}}} (\omega) \, d\omega \, ,
\end{equation}
where $\rho_{{\mathbb D}_{\mathbf{k}}}(\omega)$ and $\rho^{dis}_{{\mathbb D}_{\mathbf{k}}}(0)$ are respectively as per Equations \eqref{eq: spectral density} and \eqref{eq: residue}.
Eventually the sought advanced-minus-retarded Green's operator reads setting $x=(t,\mathbf{x},z)$,
\begin{equation}\label{eq: full adv-ret}
    \mathcal G(x,x') = \int_{\mathbb R^2} \frac{\, d\mathbf k}{2\pi} \, e^{i[\omega(t-t')-\mathbf k\cdot (\mathbf x-\mathbf x')]}  \Big\{ \rho^{dis}_\mathbb D(0) + \int_{|\mathbf k|}^{+\infty} d\omega \,[\rho_\mathbb D(\omega) + \rho_\mathbb{D}(-\omega)] \Big\}\Upsilon^0 \, .
\end{equation}

\begin{remark}\label{Rem: Pole Occurrence}
	In order to infer in which cases it is necessary to take into account the existence of a pole at $\omega=0$ in the construction of the advanced and retarded propagators, one should first focus on Equation \eqref{eq: pads system dirac eq}. Setting $\omega=0$, the square-integrable solutions can be constructed out of the building blocks
	\begin{equation}
	\varphi_1(z) = \sqrt{z}\begin{bmatrix}
		       K_{\nu}(|\mathbf k|z) \\
		      -\frac{ik_1-k_2}{|\mathbf k|}K_{\nu-1}(|\mathbf k|z) \\
		      0 \\
		      0
		    \end{bmatrix} \, , \quad 
	    \varphi_2(z) = \sqrt{z}\begin{bmatrix}
		       0 \\
		       0 \\
		       K_{\nu-1}(|\mathbf k|z) \\
		      -\frac{ik_1-k_2}{|\mathbf k|}K_\nu(|\mathbf k|z)
		    \end{bmatrix} \, .
	\end{equation}
	Their rescaled boundary values as per Equation \eqref{eq: rescaled spinors} are 
	\begin{equation}\label{eq: rescaled bdry values square integrable modes}
	_R\varphi_1(0) = \begin{bmatrix}
		       2^{\nu-1} |\mathbf k|^{-\nu}\Gamma(\nu) \\
		      -\frac{ik_1-k_2}{|\mathbf k|} 2^{-\nu} |\mathbf k |^{\nu-1}\Gamma(1-\nu) \\
		      0 \\
		      0
		    \end{bmatrix} \, , \quad 
	    _R\varphi_2(0) = \begin{bmatrix}
		       0 \\
		       0 \\
		       2^{-\nu} |\mathbf k |^{\nu-1}\Gamma(1-\nu) \\
		      -\frac{ik_1-k_2}{|\mathbf k|}2^{\nu-1} |\mathbf k|^{-\nu}\Gamma(\nu)
		    \end{bmatrix} \, .
	\end{equation}
	Hence, every linear combination $\varphi_{L^2}(z) = a\,\varphi_1(z) + b\,\varphi_2(z)$, with $a,b\in\mathbb C$, has to fulfill
	\begin{equation}\label{eq: square integrable bdry condition}
	    _R\varphi_{L^2}^{(1)}(0) =  M\, _R\varphi_{L^2}^{(2)}(0) \, ,
	\end{equation}
	where $ M= \left(\begin{smallmatrix}
		        c_1 & c_2 \\
		        c_3 & c_4
		    \end{smallmatrix}\right)\in\mathcal{M}(\mathbb{C},2)$ is such that
	\begin{equation*}
	        \left(\frac{|\mathbf k|}{2}\right)^{2\nu-1}\frac{\Gamma(1-\nu)}{\Gamma(\nu)}c_1 - \frac{ik_1-k_2}{|\mathbf k|}c_2 = 
	        \frac{ik_1+k_2}{|\mathbf k|}c_3+\left(\frac{|\mathbf k|}{2}\right)^{1-2\nu}\frac{\Gamma(\nu)}{\Gamma(1-\nu)}c_4  \, .
	\end{equation*}
	The example detailed above abides by Equation \eqref{eq: square integrable bdry condition} with $M=0$.
\end{remark}

\paragraph{Proper MIT-Bag condition --}  We consider now the MIT-Bag conditions as per Remark \ref{remark: MIT-Bag}, namely $\Upsilon^1 \,_R\Psi(0) = \,_R\Psi(0)$ where we are considering once more the rescaled spinors. Splitting $\Psi = \Psi^{(1)}\oplus\Psi^{(2)}$ allows us to rewrite this condition as 
\begin{gather*}
    \Upsilon^1 \,_R\Psi(0) = \,_R\Psi(0) \, , \quad  \Leftrightarrow  \quad \begin{cases}
        \hat\gamma^0 \,_R\Psi^{(1)}(0) = -\,_R\Psi^{(1)}(0) \, , \\
        \hat\gamma^0 \,_R\Psi^{(2)}(0) = \,_R\Psi^{(2)}(0) \, ,
    \end{cases} \\ \\ \Leftrightarrow 
    (\mathbf{1} + \hat\gamma^0)\,_R\Psi^{(1)}(0) -i (\mathbf 1 -\hat{\gamma}^0)\,_R\Psi^{(2)}(0) = 0 \, ,
\end{gather*}
which, comparing with Table \ref{tab: boundary conditions pads}, corresponds to $U = -\hat\gamma^0$. The procedure is identical to that of the previous case and, therefore, we limit ourselves to highlight that comparing this boundary condition with Equation \eqref{eq: square integrable bdry condition} entails that the pole at $\omega=0$ does not occur. In addition the eigenfunctions that fulfill this boundary condition are  $\{w_1,w_2\}$, the first two columns of $\mathscr F_\omega(z)$ as per Equation \eqref{eq: fundamental matrix}.  In this case the advanced-minus-retarded Green's operator reads, setting $x=(t,\mathbf{x},z)$,
\begin{equation}\label{eq: mit full adv-ret}
    \mathcal G(x,x^\prime) = \int_{\mathbb R^2} \frac{\, d\mathbf k}{2\pi} \, e^{i[\omega(t-t')-\mathbf k\cdot (\mathbf x-\mathbf x')]}    \int_\mathbb{R} d\omega \,[\rho_{{\mathbb D}_{\mathbf{k}}}(\omega) + \rho_{{\mathbb D}_{\mathbf{k}}}(-\omega)] \Upsilon^0 \, ,
\end{equation}
where 
\begin{equation*}
    \rho_{{\mathbb D}_{\mathbf{k}}}(\omega) = \Theta(\omega^2-|\mathbf{k}|^2)\left(w_1(z)w_1(z')^\dagger + w_2(z)w_2(z')^\dagger\right).
\end{equation*}

Following the same rationale of \cite{dappiaggi1}, the absence of a point spectrum for $\mathbb{D}_{\mathbf{k}}$ codified in the pole at $\omega=0$ entails that we can use Equation \eqref{eq: mit full adv-ret} to construct the 2-point correlation function of the ground state of a Dirac fields. More precisely we look for two bi-distributions $\omega_2^\pm$ such that their integral kernel is a solution of
\begin{equation}
\label{eq: dirac 2-pt equation of motion}
    D_x\,\omega_2^+(x,y) = 0 \, , \quad D_y\,\omega_2^-(x,y) = 0 \, ,
\end{equation}
where $D$ is the Dirac operator and where we also impose the proper MIT-bag boundary conditions. In addition we require the canonical anticommutation relations to hold true, namely
\begin{equation}
	\label{eq: dirac 2-pt CAR}
	\omega_2^+(x,y) + \omega_2^-(y,x) = i\mathcal G(x,y),
\end{equation} 
where $\mathcal G(x,y)$ is as per Equation \eqref{eq: full adv-ret}. A concrete realization of these conditions is
\begin{align}
\begin{split}\label{eq: 2-pt function to compare with allen}
    &\omega_2^+(x,x') = \int_{\mathbb R^2} \frac{\, d\mathbf k}{2\pi} \, e^{i[\omega(t-t')-\mathbf k\cdot (\mathbf x-\mathbf x')]} \int_{|\mathbf k|}^{+\infty} d\omega \, [w_1(z)\overline{w}_1(z') + w_2(z)\overline{w}_2(z') ]  \, , \\
   & \omega_2^-(x,x') = \mathcal G (x',x) - \omega_2^+(x',x) \, ,
   \end{split}
\end{align}

\begin{remark}
	We highlight that the two-point function for a Dirac field abiding by the proper MIT-bag boundary condition has been also discussed in \cite{allen} and a direct analysis of this reference compared to ours shows that the two constructions yield the same outcome. 
\end{remark}

We conclude this Section by stressing that Equation \eqref{eq: 2-pt function to compare with allen} identifies a ground state for the Dirac field subject to the proper MIT-bag boundary conditions as one can infer either by observing that only positive frequencies occur in the mode decomposition or by employing the definition given in \cite{verch}. At last, we remark that Equation \eqref{eq: 2-pt function to compare with allen} fits with the construction discussed in \cite{wrochna}. Therein it is proven that, when considering the Dirac equation on any manifold with bounded geometry, a two-point function for the Dirac field for which $\omega_2^+$ is built out of only positive-energy solutions, while $\omega_2^-$ is built out of only negative-energy solutions, determine always a quasifree, Hadamard, ground state.

\section{Conclusions}\label{Sec: Conclusions}

In this paper we have investigated massive Dirac fields on AdS$_n$, $n\geq 2$, both in the global and in the Poincar\'e patch. By generalizing a procedure first outlined in \cite{blanco} in the two-dimensional scenario, we have classified the admissible boundary conditions which can be assigned at the conformal boundary. Among them we have individuated those going under the name of MIT-bag in the literature as well as a much larger class which we have tagged as generalized MIT-bag boundary conditions. All these options entail the well-posedness of the mixed initial-boundary value problem as well as the existence of two natural candidates for playing the roles of advanced and retarded propagators for the Dirac operator. In turn, these allow to codify the canonical anti-commutation relations and to construct $\omega_2$, the two-point correlation function of an underlying ground state with prescribed boundary conditions, provided that no bound state modes are present, a feature possible when considering some among the generalized MIT-bag boundary conditions. We have applied this paradigm to two concrete examples in PAdS$_4$, but the procedure carries no a priori limitation and it could be applied in all dimensions. If one considers any globally hyperbolic open region of the underlying background, by using \cite{verch}, we can conclude that $\omega_2$ is of local Hadamard form. Hence this is a natural candidate to construct a local and covariant algebra of Wick polynomials, the building block to deal with interactions at a perturbative level. In addition, in sharp contrast with the scalar case, it appears that, when bound state modes are absent, no infrared singularity appears, hence avoiding the potential pathological features first highlighted in \cite{Campos:2023mwh}.

We reckon that this work represents just the first step in the analysis of the interplay between boundary conditions and Dirac fields. There are several open questions which should be tackled in future works and, among these, we highlight the following:
\begin{itemize}
	\item From a structural viewpoint, the advanced and retarded propagators $\mathcal{G}^\pm$, constructed in Section \ref{sec: green operator dirac equation} abide by all the defining properties of a fundamental solution, but it is not manifest that, given $f\in C^\infty_0(\mathbb{R}^+\times\mathbb{R}^{n-1};\mathbb{C}^{2^{\lfloor\frac{n}{2}\rfloor}})$,
	$$\textrm{supp}(\mathcal{G}^\pm(f))\subseteq J^\mp(\textrm{supp}(f)).$$
	In globally hyperbolic spacetime this feature is codified directly in the definition of the advanced and retarded propagators, playing a key role in proving their uniqueness. When one is working with a mode decomposition this property is not manifest, but it can be proven using energy estimates. It has been highlighted that, even in the scalar case, this approach does not work and only recently an alternative procedure has devised in \cite{Costeri:2025fqn}. Yet, since in this reference, it is particularly important that the dynamics is codified by a second order partial differential equation, a direct application to Dirac fields is not immediate. Hence it would be desirable to devise a workaround which allows to prove the support property also for the propagators of a Dirac field,
	\item Still from the structural viewpoint, one of the pillars of the algebraic approach to quantum field theory goes under the name of Radzikowski theorem \cite{Radzikowski:1996pa, Radzikowski:1996ei}. This guarantees the equivalence between the local and global Hadamard form of a two-point correlation function. The first establishes the existence of a codified form of the integral kernel $\omega_2(x,x^\prime)$ provided that $x$ and $x^\prime$ lie in the same geodesic neighbourhood. This guarantees that the singular part of $\omega_2$ is covariant, namely it is completely determined in terms of the underlying geometry and of the equation of motion. At the same time the global Hadamard form codifies the singular structure of $\omega_2$ globally using the language of microlocal analysis. The interplay and equivalence of these two conditions is foundational for establishing a perturbative, local and covariant quantization scheme for interacting field theories, see {\it e.g.} \cite{Rejzner}. Yet all these results rest on the assumption of global hyperbolicity and, in presence of timelike boundaries such as in AdS$_n$, more sophisticated techniques need to be employed. In the scalar case, the global structure of the two-point function has been studied in \cite{GaWr18, Dappiaggi-Marta_2020} and a generalization of Radzikowski theorem has been proven in \cite{Costeri:2025fqn}. Yet all these works strongly rely on the underlying equation being of second order, hence requiring a substantial modification to be applied to Dirac fields. Furthermore, in \cite{DGM} Hadamard states with MIT boundary conditions have been studied and a comparison with our results is certainly desirable.
	\item From a physical perspective, we would like to draw the attention on two open questions. In view of finding a connection with the AdS/CFT correspondence, one should explore the possibility of considering a larger class of boundary conditions, in which the bulk field restricted to the boundary acts a source of a dynamics thereon. In the scalar scenario, dynamical boundary conditions of Wentzell type have been already investigated, see, {\it e.g.}, \cite{dappiaggi-benito1, dappiaggi-benito2, Juarez-Aubry:2020aoo}. Yet, it is unclear whether a similar option exists also when considering field theories whose dynamics is codified by Dirac type operators and a detailed analysis should be undertaken. A second remarkable and recent physical application explores the interplay between quantum field theory and relativistic quantum information by coupling a field to an Unruh-de Witt detector, which is a two-level system moving along a worldline of the underlying background. The possibility of having different boundary conditions has far reaching consequences which have been investigated when considering scalar fields, even in connection to specific phenomena such as the anti-Unruh and anti-Hawking effects, see, {\it e.g.}, \cite{Hodgkinson:2012mr,Ng:2014kha,Henderson:2019uqo,DeSouzaCampos:2020ddx,deSouzaCampos:2021awm,Ramos:2025xyg}. Since Dirac fields admit boundary conditions which are structurally very different from the scalar counterparts, one can expect qualitative and quantitative differences which we deem worth to be investigated. 
\end{itemize}

\section*{Acknowledgments}

C.D. is grateful to the GNFM--Indam (Gruppo Nazionale di Fisica Matematica) for the support in the realization of part of this work. The work of A.P. is supported by a PhD fellowship of the University of Pavia.

\appendix

\section{The Dirac Equation on Curved Backgrounds}\label{Sec: Appendix A}

Goal of this appendix is to recall succinctly the geometric construction of spinors on $(\mathcal M, g)$, an $n$-dimensional Lorentzian, orientable, time-orientable, smooth manifold endowed with a metric $g$, where $\mathcal M$ has possibly a non empty boundary, see also \cite{spin}.  By $F\mathcal{M}$ we denote the frame bundle associated to $\mathcal M$ while $\varepsilon\in\Gamma(F\mathcal M)$ is a smooth section, whose components $\{\varepsilon_a(p)\}_{a=0,\dots,n-1}$ identify an ordered, orthonormal, orthochronous basis for $T_p\mathcal M$. In addition, if we fix $\{e_\mu\}_{\mu=0,\dots,n-1}$, a local coordinate system for $\mathcal M$, we say that $\varepsilon$ is a \textbf{vielbein} if 
\begin{equation}
	\label{eq: vielbein diagonalization}
	g^{\mu\nu} = -\eta^{ab}(\varepsilon_a)^\mu (\varepsilon_b)^\nu, \quad \forall \, a,b,\mu,\nu = 0,\dots,n-1,
\end{equation}
where $g_{\mu\nu}$ and $\eta_{ab}$ denote the components respectively of the metric tensor of $\mathcal M$ and of Minkowski spacetime with respect to the bases $\{e_\mu\}$ and $\{\varepsilon_a\}$, while $(\varepsilon_a)^\mu$ are such that $\varepsilon_a=\varepsilon_a^\mu e_\mu$. On top of these we also require $\mathcal{M}$ to admit a spin structure $(S\mathcal M,\pi)$ where $S\mathcal M$ is the spin bundle, that is a principal bundle whose structure group is $Spin_0(1,n-1)$, the double cover of $SO_0(1,n-1)$, the component connected to the identity of the Lorentz group, while $\pi:S\mathcal M\to F\mathcal M$ is a morphism of principal bundles covering the identity. Existence of such structure is tantamount to requiring that the second Stiefel-Withney class $w_2(M)\in H_2(M;\mathbb{Z}_2)$ vanishes, while the first one $w_1(M)\in H_1(M;\mathbb{Z}_2)$ classifies the number of non equivalent choices. In this work we always consider manifolds for which both $w_1(M)$ and $w_2(M)$ vanish. In addition to these data, we consider on top of Minkowski spacetime $(\mathbb{R}^n,\eta)$ the standard algebra of gamma-matrices $\gamma^a$, $a=0,\dots,n-1$ such that $\{\gamma^a,\gamma^b\}=2\eta^{ab}\mathbf{1}_{\mathbb{C}^{2^{\lfloor\frac{n}{2}\rfloor}}}$ where $\mathbf{1}$ denotes the identity matrix acting on the spinor representation space $\mathbb{C}^{2^{\lfloor\frac{n}{2}\rfloor}}$. These can be used to define a counterpart on $(\mathcal M,g)$ by means of the defining identity
	\begin{equation}
	\label{eq: covariant gamma matrices}
	\gamma^\mu := \gamma^a(\varepsilon_a)^\mu, \quad a=0,\dots,n-1,
\end{equation}
which entails in turn
\begin{equation}
	\label{eq: curved space clifford algebra}
	\{\gamma^\mu,\gamma^\nu\}(x) := \gamma^\mu(x)\gamma^\nu(x)+ \gamma^\nu(x)\gamma^\mu(x) = 2g^{\mu\nu}(x)\mathbf1_{\mathbb{C}^{2^{\lfloor\frac{n}{2}\rfloor}}}.
\end{equation}

Starting from these premises we consider on top of $\mathcal M$ the trivial bundle $\mathcal{D}(\mathcal{M})\equiv\mathcal M\times\mathbb{C}^{2^{\lfloor\frac{n}{2}\rfloor}}$ and we denote its smooth sections as {\bf Dirac fields/spinors} $\psi:\mathcal M\to\mathbb{C}^{2^{\lfloor\frac{n}{2}\rfloor}}$. On top of $\Gamma(D\mathcal M)\equiv C^\infty(\mathcal M;\mathbb{C}^{2^{\lfloor\frac{n}{2}\rfloor}})$ we define the spin connection 
$$\nabla: \Gamma(T\mathcal M) \times \Gamma(D\mathcal M) \to \Gamma(D\mathcal M),$$ 
such that, for all $(X,\psi)\in\Gamma(T\mathcal M) \times \Gamma(D\mathcal M)$, 
\begin{align}
	\label{eq: spinor covariant derivative}
	\begin{split}
		\nabla_X\psi &= X^\mu\partial_\mu\psi + \frac18 X^\mu[(\varepsilon_a)_\nu\partial_\mu (\varepsilon_b)^\nu + \Gamma_{\mu\nu}^\rho (\varepsilon_a)_{\rho}(\varepsilon_b)^\nu][\gamma^a,\gamma^b] \psi= \\
		& = X^\mu\partial_\mu\psi + \frac18 X^\mu\{[\gamma_\nu,\partial_\mu\gamma^\nu] + \Gamma^\rho_{\mu\nu}[\gamma_\rho,\gamma^\nu]\} \psi,
	\end{split}
\end{align}
where $\Gamma_{\mu\nu}^\rho$ denote the Christoffel symbols induced by $g$ while $[\cdot , \cdot]$ denotes the matrix commutator. Consequently we can define the {\bf (massless) Dirac operator}
\begin{equation}
	\label{Eq: dirac operator}
	D:\Gamma(D\mathcal M) \to \Gamma(D\mathcal M), \quad D\psi = \gamma^\mu\nabla_\mu\psi,
\end{equation}
which abides by the Lichnerowicz–Weitzenb\"ock formula, namely
	\begin{equation}\label{Eq: weitzenbock formula}
	DD^*\psi = D^*D\psi = \square_g\psi + \frac{1}{4}R\psi,
\end{equation}
where $D^*$ is the formal adjoint of $D$, and where $\Box_g$ and $R$ are respectively the D'Alemebert wave operator and the scalar curvature built out of $g$. To conclude the section we introduce the dual Dirac bundle $\mathcal{D}^*\mathcal{M}\equiv\mathcal M\times(\mathbb{C}^{2^{\lfloor\frac{n}{2}\rfloor}})^*$ whose smooth sections are called {\bf Dirac cospinors} $\widetilde{\psi}:\mathcal M\to (\mathbb{C}^{2^{\lfloor\frac{n}{2}\rfloor}})^*$.  We can relate spinor and cospinors via a \textit{conjugation}
\begin{equation}
	\label{eq: dirac conjugation}
	\overline{\cdot} : \Gamma(D\mathcal{M}) \to \Gamma(D^*\mathcal M), \quad \overline\psi = \psi^\dagger\gamma^0,
\end{equation}
where $\psi^\dagger$ is the adjoint of $\psi$, while the index of $\gamma^0$ is intended as $a=0$ and not $\mu=0$, see Equation \eqref{eq: covariant gamma matrices}. Notice that the Dirac equation $(D-m)\psi = 0$ implies the anti-Dirac equation
\begin{equation}\label{eq: anti-dirac equation}
\overline{\psi} (\overline D +m) = 0 \, ,
\end{equation}
where we implicitly assume that $\overline D + m$ acts on the left, while
\begin{align}
\overline \psi \, \overline D = -\overline{D\psi}  = \partial_\mu\overline \psi \gamma^\mu - \frac18 \overline \psi \gamma^\mu\{[\gamma_\nu,\partial_\mu\gamma^\nu] + \Gamma^\rho_{\mu\nu}[\gamma_\rho,\gamma^\nu]\}  \, .
\end{align}

\noindent To conclude this appendix, we highlight that, in this work, we make the following choices of $\gamma$-matrices which are elements of $\mathcal{M}(2^{\lfloor \frac{n}{2}\rfloor};\mathbb{C})$:
\begin{itemize}
	\item For $n=2$, $\gamma^0 = -i\big( \begin{smallmatrix}
		1 & 0 \\
		0 & -1
	\end{smallmatrix} \big)$ , $\gamma^1 = \big( \begin{smallmatrix}
		0 & 1 \\
		1 & 0
	\end{smallmatrix} \big)$ .
	
	\item  For $n=3$, in addition to $\gamma^0$ and $\gamma^1$ as above, we set $\gamma^2 = \big( \begin{smallmatrix}
		0 & -i \\
		i & 0
	\end{smallmatrix} \big)$ .
	
	\item For $n>2$ even, denoting by $\hat\gamma^a$ the gamma-matrices in the $(n-1)$-dimensional case 
	\begin{equation*}
		\gamma^0 = -i \begin{pmatrix}
			\mathsf{1}_{2^{\lfloor \frac{n}{2}\rfloor-1}} & 0 \\
			0 & -\mathsf{1}_{2^{\lfloor \frac{n}{2}\rfloor-1}}
		\end{pmatrix} \, ,\quad \gamma^1 = \begin{pmatrix}
		0 & \hat\gamma^{0} \\
		-\hat\gamma^{0} & 0 
		\end{pmatrix}, \quad \gamma^a = \begin{pmatrix}
			0 & i\hat\gamma^{a-1} \\
			-i\hat\gamma^{a-1} & 0 
		\end{pmatrix} \, , 
	\end{equation*}
	where $a=1,\dots,n-1$,  $1_{\frac n2}$ is the $\frac n2 \times \frac n2$ identity, while the if $a>1$ while $\hat{\gamma}^0=i\gamma^0$.
	
	\item For $n>3$ odd,
	\begin{equation*}
		\gamma^0=i\hat{\gamma}^0,\quad\gamma^a = \hat\gamma^a \, , \ a=1,\dots,n-2 \, , \quad \gamma^{n-1} = \begin{pmatrix}
			0 & \mathsf{1}_{2^{\lfloor \frac{n}{2}\rfloor-1}} \\
			\mathsf{1}_{2^{\lfloor \frac{n}{2}\rfloor-1}} & 0 
		\end{pmatrix} \, .
	\end{equation*}
\end{itemize}

\end{document}